\documentclass[twocolumn,superscriptaddress]{revtex4-2}

\usepackage[usenames,dvipsnames]{xcolor}

\usepackage{lettrine}

\usepackage{amsfonts, amsmath, amssymb}
\usepackage[utf8]{inputenc}
\usepackage[english]{babel}
\usepackage{graphicx}
\usepackage{braket}
\usepackage{wrapfig}
\usepackage{hyperref}
\usepackage{bm}
\usepackage{color}
\usepackage{soul}
\usepackage{mathtools}
\usepackage{float}
\usepackage{ulem}
\usepackage{cancel}
\newcommand{\bb}[1]{\mathbf{#1}}
\newcommand{\m}[1]{\mathcal{#1}}

  {\left\lbrace\begin{array}{@{}l@{}}}%
  {\end{array}\right.}

\newcommand{\blank}{\vspace{3mm}\noindent}

\begin{document}

\title{Kondo-Zeno crossover in the dynamics of a monitored quantum dot}

\author{Matthieu Vanhoecke}
\email{matthieu.vanhoecke@college-de-france.fr}
\affiliation{JEIP, UAR 3573 CNRS, Coll\`ege de France,   PSL  Research  University, 11,  place  Marcelin  Berthelot,75231 Paris Cedex 05, France}
\author{Marco Schir\`o}
\email{marco.schiro@college-de-france.fr}
\affiliation{JEIP, UAR 3573 CNRS, Coll\`ege de France,   PSL  Research  University, 11,  place  Marcelin  Berthelot,75231 Paris Cedex 05, France}

\begin{abstract}
Continuously monitoring a quantum system can strongly affect its properties and even suppress its coherent evolution via the Quantum Zeno effect.
Well understood for few body  quantum systems, the role of quantum measurements on entangled many-body states is still largely unexplored. Here we focus on one of the simplest entangled many-body state, arising via the Kondo effect in a strongly interacting quantum dot coupled to a metallic bath, and investigate the effect of continuous monitoring of the dot total charge.  We show that the decay rate of an initially polarized spin displays a crossover from Kondo screening, with a decay rate controlled by interactions, to Quantum Zeno effect, with a decay rate which decreases with bare dissipation as the monitoring rate is increased.  Remarkably we show that the long-lived Kondo state is robust to weak dissipation, as further confirmed by the dot spectral function which features a clear Kondo peak at finite dissipation, even in a regime where charge fluctuations and the associated Hubbard bands have been quenched by the monitoring protocol. We derive an effective model for the long-time dynamics which is described, at weak dissipation, by a non-Hermitian Kondo model with complex-valued spin exchange which is known to host exotic low-energy physics and a dissipative phase transition between Kondo and non-Kondo steady-state. Finally,  as the dephasing is increased heating due to doublon production takes over and control the spin decay. 
\clearpage
\end{abstract}
\maketitle

\lettrine{A}{} spinful quantum dot coupled to a metallic bath represents one of the simplest many-body problem. Strong correlations in the dot freeze charge fluctuations and leave a  fluctuating quantum spin which is collectively screened at low energy into an entangled Kondo singlet state~\cite{anderson1961localized,kondo1964resistance,nozieres1974fermiliquid,wilson1975therenormalization,hewson1993thekondo} which leaves behind
unique equilibrium~\cite{mitchell2016kondo,shim2023hierarchical} and dynamical signatures, including a ultra slow spin dynamics on the scale of the inverse Kondo temperature~\cite{nordlander1999how,anders2005realtime}. First observed in metals with diluted magnetic impurities, the Kondo effect has been later realized in quantum dots and various mesoscopic setups~\cite{goldhaber1998kondo,Kouwenhoven_2001,Pustilnik_2004,latta2011quantum,kurzmann2021kondo,piquard2023observing}.
A different dissipative mechanism to freeze or slow down the dynamics of a system is the continuous Quantum Zeno Effect~\cite{misra1977a,facchi2002quantum}, where projective measurements or strong continuous monitoring leads to a complete localization of the dynamics. The Zeno Effect has been observed in different quantum platforms, from cavity and circuit QED~\cite{signoles2014confined,bretheau2015quantum} to ultracold atoms
\cite{garcia-ripoll2009,schafer2014experimental,meausurement2015patil} and nuclear spins~\cite{kalb2016experimental}, and has recently raised new interest in many-body settings~\cite{li2018quantum,Froml2019,Biella2021manybodyquantumzeno}.

Recent experimental progress in quantum simulation platforms both in the solid state and with ultracold atoms have opened new 
windows to explore the physics of dissipative systems and offer now the possibility to explore the interplay between Kondo screening and Zeno effect. This is the case of quantum dots coupled to a quantum point contact acting as a monitoring device~\cite{avinun2004controlled,kang2007entanglement,aono2008dephasing,sukhorukov2007conditional,ferguson2023measuremnt,wiseman2009quantummeasurementand,sankar2024detectortuned} or of ultracold alkaline-earth atoms where Kondo physics has been realized~\cite{riegger2018localized,zhang2020controlling,nagy2018exploring}, which are naturally exposed to correlated dissipative processes, such as dephasing due to spontaneous emission~\cite{gerbier2010heating,bouganne2020} or two-body losses due to inelastic scattering~\cite{garcia-ripoll2009,TomitaEtAlScienceAdv17,honda2022observation}. 
The physics of these dissipative quantum impurity models has started to be explored only recently, with a focus on non-interacting chains with localized single particle losses~\cite{Froml2019,damanet2019controlling,visuri2022symmetry,visuri2023nonlinear,visuri2023dc} or pumps~\cite{krapivsky2019free,krapivsky2020free} or local dephasing~\cite{Scarlatella2019,tonielli2019orthogonality,dolgirev2020nongaussian,Chaudhari_2022,ferreira2023exact}. Non-Hermitian quantum impurity models, arising from a postselection over quantum trajectories, have also been studied~\cite{Nakagawa2018,
yoshimura2020nonhermitian,stefanini2023orthogonality}. Recently the effect of projective measurements on the Kondo effect in the steady-state have been discussed~\cite{hasegawa2021kondo}.
In addition to their intrinsic interest, dissipative quantum impurity models also arise as effective description of open Markovian lattice models in the large connectivity limit, within Dynamical Mean-Field Theory~\cite{scarlatella2021dynamical}.

\begin{figure}[!t] 
 \includegraphics[width=0.4\textwidth]{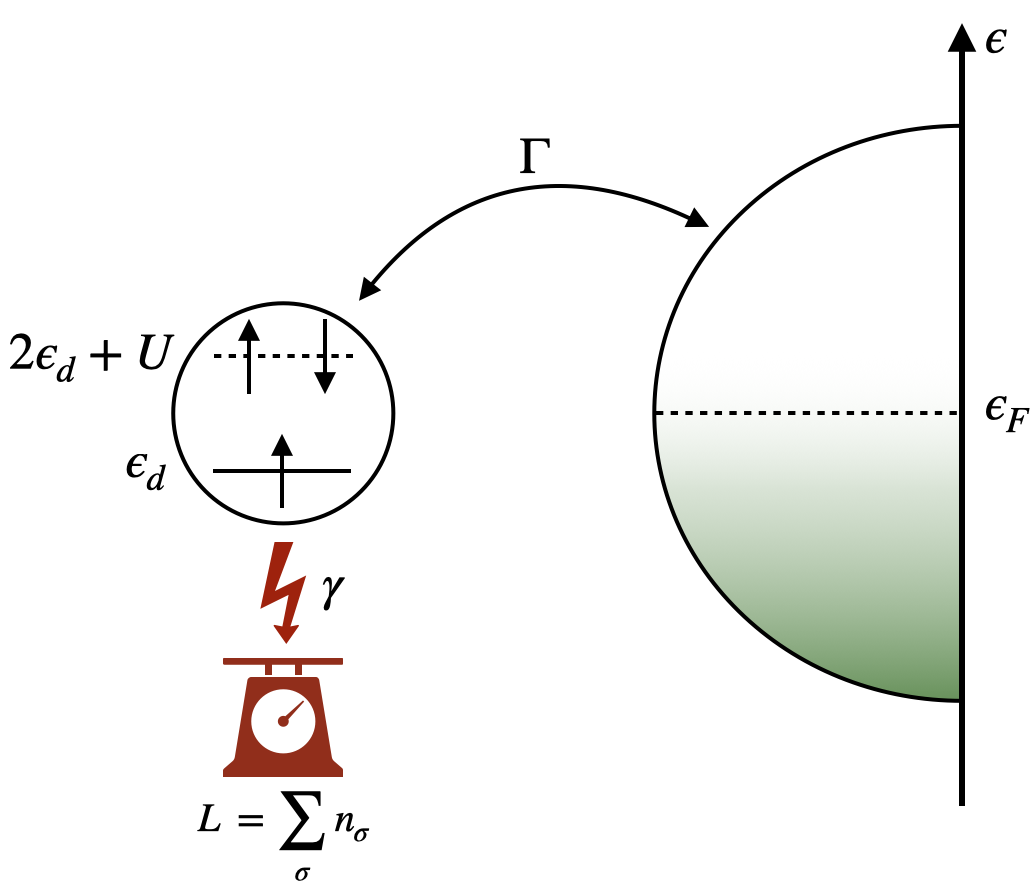}
    \caption{\label{fig:sketch} Sketch of the setup for the Kondo-Zeno Crossover.  A singly-occupied quantum dot is coupled to a large metallic bath via a hybridization $\Gamma$ and to a measurement apparatus continuously monitoring, with rate $\gamma$,  its total charge.  Upon averaging over the monitoring noise the dynamics of the dot plus metallic bath is described by Lindblad master equation with a jump operator $L=\sum_{\sigma}n_{\sigma}$, describing charge dephasing. }
\end{figure}

In this work we investigate the effect of continuous monitoring on the dynamics of Kondo effect. Specifically we study the time evolution of an interacting quantum dot coupled to a metallic bath, in presence of additional dissipation (dephasing) due to monitoring of quantum dot's total charge (see Fig.~\ref{fig:sketch}). Since both Coulomb repulsion and charge monitoring compete to freeze charge fluctuations without affecting the spin degrees of freedom one can expect a non-trivial effect on Kondo physics.  We show that the dynamics of an initially polarised spin displays a crossover from Kondo to Zeno screening, as dissipation is increased, as highlighted by a non-monotonous spin-relaxation rate as a function of dissipation. Using a Schrieffer-Wolff transformation, we show that the effective model controlling this crossover takes the form of a non-Hermitian Kondo model, with complex (i.e. coherent and dissipative) Kondo coupling and in presence of additional dissipative terms describing doublon production. The former controls the physics at weak monitoring and leads to a robust Kondo effect even in presence of dissipation,  as we further confirm by computing the spectral function and the impurity entanglement,  while the latter dominate at strong dephasing and give rise to local heating.

\blank
\textbf{Results }\\
{\bf Dissipative Anderson Impurity Model }

%%%%%%%%%%%%%%%%%%%%%%%%%%%%%%%%%%%%%%%%%%%%%%%%%%%%%%%%
We consider a model for an interacting spinful quantum dot coupled to a large metallic bath and to a monitoring device which measures weakly but continuously its total charge (see Fig.~\ref{fig:sketch}),  as realized in the homodyne detection or the quantum-state diffusion protocol~\cite{gisin1992thequantumstate,wiseman2009quantummeasurementand}.
%%%%%%%%%%%%%%%%%%%%%%%%%%%%%%%%%%%%%%%%%%%%%%%%%%%%%%%
\begin{figure*}[!t] 
	\centering
 \includegraphics[width=1.0\textwidth]{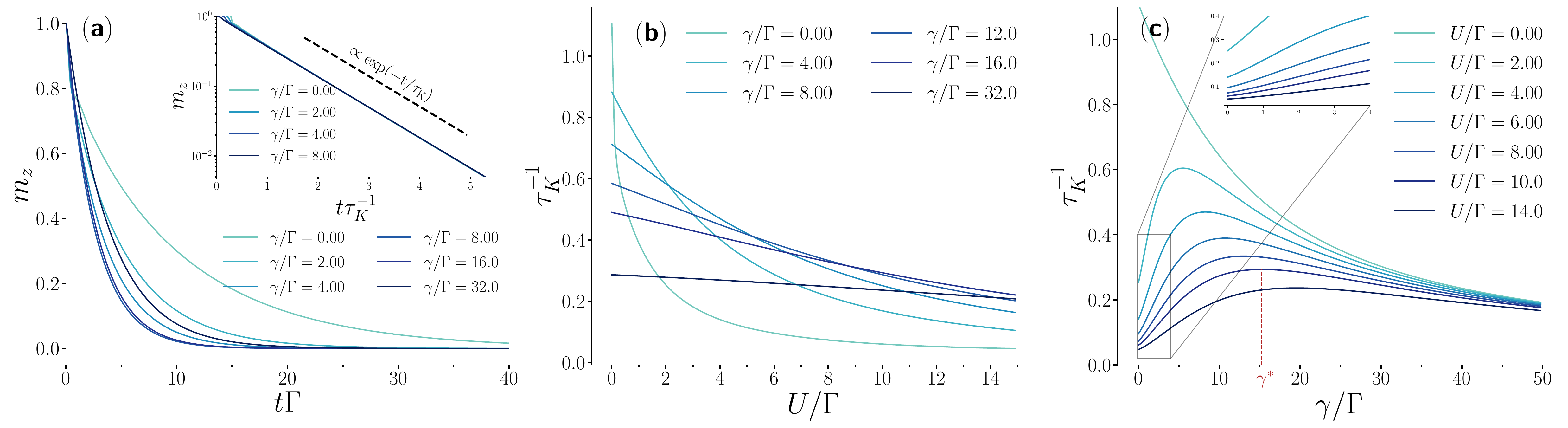}
    \caption{\label{fig:spin} Impurity Spin Dynamics and Kondo-Zeno Crossover  - (a) Dynamics of the impurity magnetization starting from a polarised state $\rho_0= \vert \uparrow \rangle \langle  \uparrow \vert $ and evolving under the  dynamics of the half-filled dissipative Anderson Impurity Model, with $U=8\Gamma$.  The impurity spin decays to zero at long times with an exponential decay $m_z\sim \exp(-t/\tau_K)$. (b) The decay rate $\tau_K^{-1}$ depends non-monotonously on the monitoring rate $\gamma$. For small dissipation the decay is faster, however upon increasing $\gamma$ we see that the decay of the impurity slows down. The crossover between these two regimes, characteristic of the continuous Quantum Zeno effect, is set by $\gamma \sim U$. (c) The crossover is also visible in the dependence of  $\tau_K^{-1}$ from the interaction $U$.}
\end{figure*}
%%%%%%%%%%%%%%%%%%%%%%%%%%%%%%%%%%%%%%%%%%%%%%%%%%%%%%%
Upon averaging over the monitoring noise the dynamics of the system  (dot plus metallic bath) is described by a dissipative Anderson impurity model with local charge dephasing for which the density matrix $\rho_t$ of the system evolves according to the many-body Lindblad master equation (Supplementary Note 1)	
\begin{align}\label{eqn:lindblad}
    \partial_t \rho_t = -i \left[ H, \rho_t \right] + \gamma\left( L \rho_t L^\dagger -\frac{1}{2} \{ L^\dagger L , \rho_t \}\right)
\end{align}
where $H$ is the dot plus bath Hamiltonian given by the Anderson Impurity Model~\cite{hewson1993thekondo}
\begin{align}\label{eq:H_aim}
H=\sum_{\bb{k},\sigma}\varepsilon_{\bb{k}}c^\dagger_{\bb{k},\sigma}c_{\bb{k},\sigma}+ \sum_{\bb{k},\sigma} \left(V_{\bb{k}} d_{\sigma}^\dagger c_{\bb{k},\sigma} + h.c \right)+H_{\rm dot}\,.
\end{align}
Here the first term describes the Hamiltonian of the metallic bath with fermionic operators $c_{\bb{k},\sigma} ,c^\dagger_{\bb{k},\sigma} $, the second term describes the hybridization between dot and bath with coupling $V_{\bb{k}}$ and  the last term describes the dot Hamiltonian 
\begin{align}
	H_{\rm dot}=\varepsilon_d \sum_{\sigma}d^\dagger_{\sigma}d_{\sigma}+Un_{\uparrow}n_{\downarrow}\,,
\end{align}
where $d^\dagger_{\sigma},d_{\sigma}$ are the dot creation/annihilation operators,  $\varepsilon_d$ is the energy level of the dot which can be controlled externally with a gate voltage, and $U$ is the Coulomb repulsion, with $n_{\sigma}=d^\dagger_{\sigma}d_{\sigma}$. The action of the monitoring device is taken into account at the Markovian level by a jump operator proportional to the dot total charge $n$, i.e. 
\begin{align}
L =L^\dagger =\sum_{\sigma}n_{\sigma}\equiv n\,,
\end{align}
where $\gamma$ in Eq.~(\ref{eqn:lindblad}) is the monitoring (or dephasing) rate~\cite{wiseman2009quantummeasurementand}. We note that this type of dissipation preserves particle-hole symmetry, namely under the transformation $n \rightarrow 1- n$ the dissipative part of the Lindblad master equation remains unaffected. As such, if we choose $\epsilon_d=-U/2$ such that the coherent evolution is also particle-hole symmetric we can conclude that the system remains half-filled under the dynamics generated by Eq.~(\ref{eqn:lindblad}).
In the following we consider a metallic bath with a semicircular density of states of bandwidth $W$ around the Fermi energy $\varepsilon_F=0$, giving rise to a hybridization function  $\Gamma(\varepsilon)= 2 \pi \sum_{\bb{k}}V_{\bb{k}}^2\delta(\varepsilon-\varepsilon_{\bb{k}})=
 \Gamma\sqrt{1-(\varepsilon/W)^2} $. We note that bath density of states does not play a major role, as long as it describes a metal with finite weight at the Fermi energy. As such different choices are possible, such as for example a flat density of states, with no changes to the qualitative physics.
 
We solve the dissipative Anderson Impurity model by means of a recently developed self-consistent dynamical map based on the Non-Crossing Approximation (NCA)~\cite{scarlatella2021dynamical,scarlatella2023selfconsistent}. This method to treat non-perturbatively both electron-electron interactions and local dissipation on the dot while treat the coupling to the metallic bath by resumming a set of diagrams in the hybridization expansion (see Methods and Supplementary Note 2). In absence of dissipation the NCA method is known to capture qualitatively the Kondo effect both in and out of equilibrium~\cite{meir1993low,nordlander1999how,rosch2001kondo}.
 
\begin{figure*}[t!]
 	\centering
    \includegraphics[width=\textwidth]{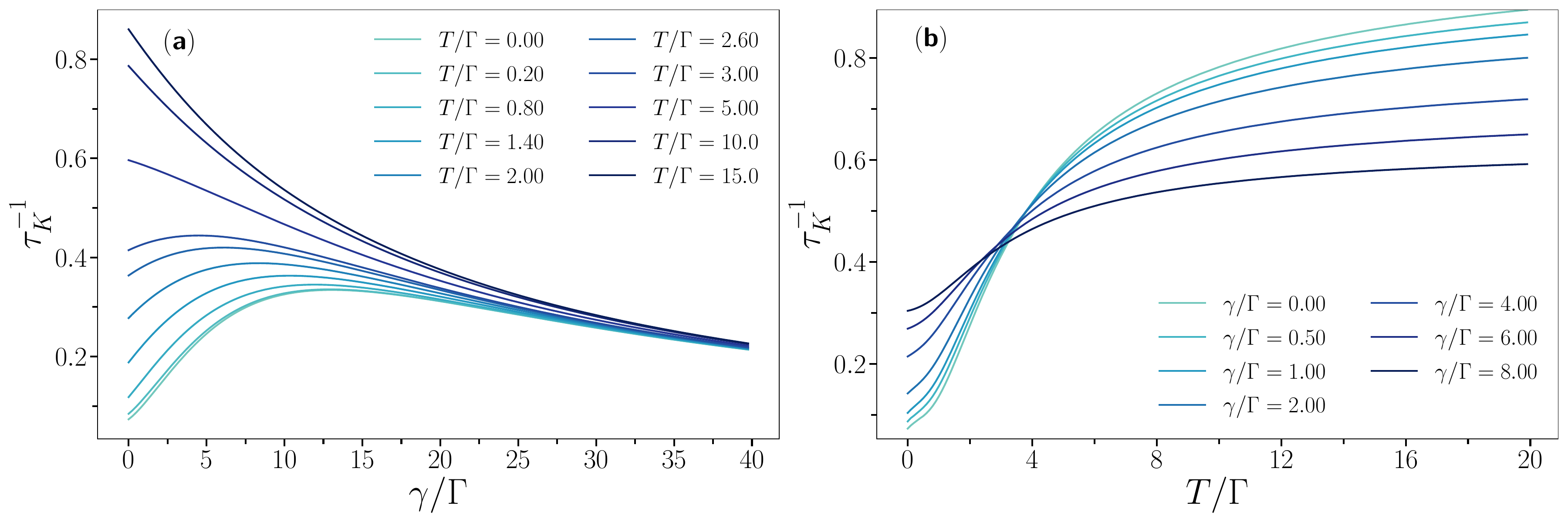}
    \caption{\label{fig:finiteT}  Dissipative Anderson Impurity Model - (a) Decay rate $\tau_K^{-1}$ as a function of the temperature $T$ of the fermionic bath, for different dissipation strenghts $\gamma$. Here we have fixed the interaction to $U=-2 \epsilon_d=8\Gamma$.  We see that the decay rate increases with temperature, while it displays a non monotonous behavior with $\gamma$, namely at low temperature the dynamics becomes faster with $\gamma$ while at high-temperature the Zeno effect brings $\tau_K^{-1}$ down as $\gamma$ increases.}
\end{figure*}

{ \bf Spin Dynamics and Kondo-Zeno Crossover}

We consider an initially polarised spin up and then suddenly switch-on the monitoring and the coupling to the metallic bath at temperature $T=0$. In Fig.~\ref{fig:spin}(a) we plot the dynamics of the dot magnetization $m_z(t)=\mbox{Tr}\rho_t \left(n_{\uparrow}- n_{\downarrow}\right)$ at fixed interaction $U=8\Gamma$ and for different values of the monitoring rate $\gamma$. We see that the magnetization decays to zero at long times for all values of $\gamma$.  Interestingly, the spin dynamics first accelerates upon adding a weak dissipation, then for large enough $\gamma$ it slows down again, see Fig.~\ref{fig:spin}(a). 
In the unitary Anderson impurity model the spin dynamics is universal as a function of $tT_K$, where $T_K$ is the Kondo temperature~\cite{anders2005realtime}.  For our dissipative case a full scaling theory of the Kondo effect is not known, yet we can show that some form of universality still holds true in the dissipative dynamics. As we show in the inset of Fig.~\ref{fig:spin} the magnetization at long times is a universal function of $t\tau_K^{-1}$, where $\tau^{-1}_K(\gamma,U)$ is the spin relaxation rate, or inverse Kondo time, extracted from an exponential fit $m_z\sim \exp(-t/\tau_K)$.  
We now discuss how this quantity depends on interaction and dissipation. In Fig.~\ref{fig:spin}(b) we plot the spin decay rate as a function of interaction $U$, for different values of $\gamma$. As mentioned earlier,  in absence of dephasing (corresponding to $\gamma=0$) the decay of the magnetization is due to the hybridization with the metallic bath leading to a Kondo singlet. The associated spin-decay rate depends strongly on interaction and it is directly related to the Kondo temperature~\cite{hewson1993thekondo,anders2005realtime,Kohn_2022,wauters2023simulations}, as we confirm with our NCA method (See Supplementary Note 3). For weak dissipation the spin decay rate keeps a non trivial $U$-dependence, suggesting the presence of a correlated dissipative many-body state, while in the strongly dissipative Zeno regime we see that the decay rate is almost independent of $U$, see Fig.~\ref{fig:spin}(b) pointing towards a different mechanism behind the spin relaxation.
The dependence of the spin relaxation rate from dissipation however contains the most intriguing result. As we see in Fig.~\ref{fig:spin}(c) the rate depends non-monotonously on $\gamma$:  it first grows at small dissipation since the system is more dissipative, then reaches a maximum at $\gamma=\gamma^*$ and decreases at large $\gamma$ as $1/\gamma$, indicating the freezing of the system dynamics due to the strong observation, a signature of the continuous Quantum Zeno ~\cite{garcia-ripoll2009,Froml2019,rossini2020b,secli2022steady,scarlatella2021dynamical}. The weak monitoring regime $\gamma<\gamma^*$ is particularly interesting as it shows a strong dependence on the interaction $U$. Indeed we see in the inset of Fig.\ref{fig:spin}(c) that in the weakly correlated regime the spin-decay rate is linear at small $\gamma$, while for $U\gg\Gamma$ a sub-linear dependence emerges (see inset) suggesting a robustness of Kondo physics to weak charge monitoring. The crossover between Kondo decay due to singlet formation and Zeno decay due to dephasing is one of the main result of this work. We note in addition that the Kondo-Zeno crossover also controls the dynamics of the impurity entanglement entropy that shows a sharp maximum at short times for dissipation smaller than $\gamma\sim U$ (See Supplementary Note 3).

%%%%%%%%%%%%%%%%%%%%%%%%%%%%%%%%%%%%%%%%%%%%%%%%%%%%%%%
 \begin{figure*}[t!]
 	\centering
 \includegraphics[width=\textwidth]{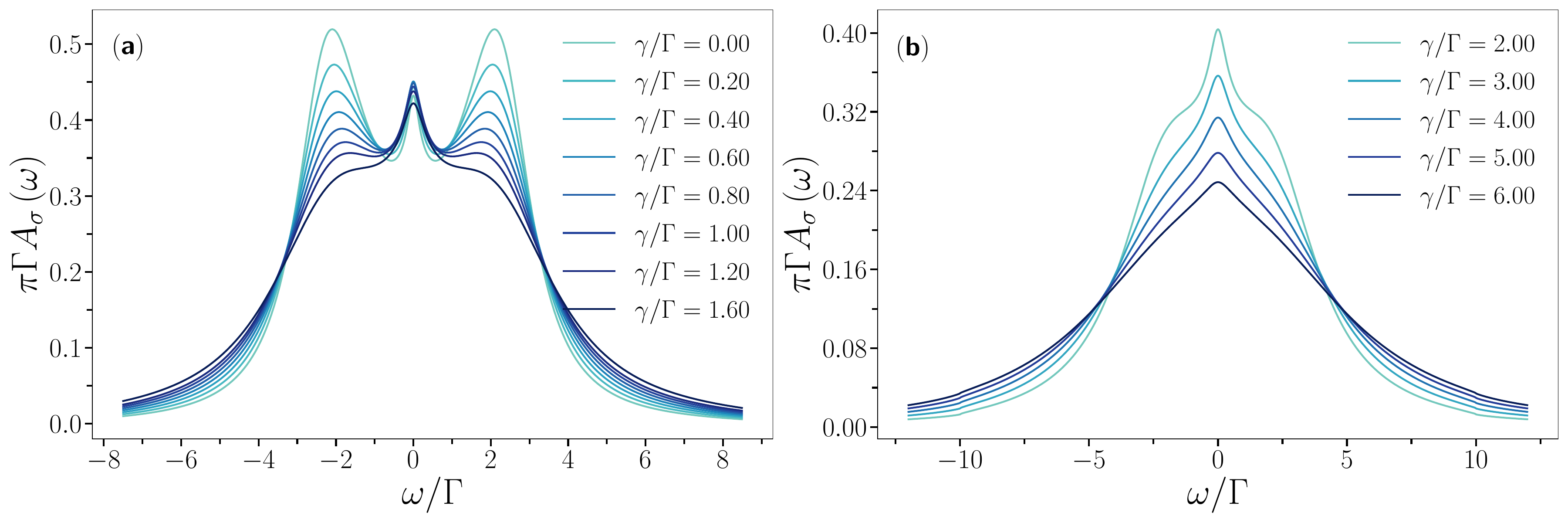}
    \caption{\label{fig:SpectralFuct2}  Dissipative Anderson Impurity Model - (a) Impurity spectral function at half-filling and for $U = -2 \epsilon_d = 4 \Gamma $, for increasing values of $\gamma$. We see that the low-frequency structure is robust to dephasing.(b) Upon increasing further $\gamma$ the central peak eventually merges with the incoherent background.}
\end{figure*}
%
% In absence of dissipation the spectrum shows a characteristic three peak structure, with coherent low-energy Kondo peak and incoherent Hubbard bands at high-frequency. As we show in Ref.~\cite{supplementary} a weak charge dephasing,  $\gamma\ll U$, strongly renormalises the latter, whose location drifts with $\gamma$, while leaves practically unaffected the Kondo peak.  
 %%%%%%%%%%%%%%%%%%%%%%%%%%%%%%%%%%%%%%%%%%%%%%%%%%%%%%%

{\bf Effect of Finite Temperature}

We now consider the effect of a finite temperature $T$ in the metallic bath on the spin relaxation and the Kondo-Zeno crossover. To this aim we compute the dynamics of the dot magnetization and extract the spin relaxation rate $\tau_K^{-1}$, that we plot it in Fig.~\ref{fig:finiteT}(a) as a function of $\gamma$ for different values of temperatures $T$, at  fixed interaction $U=8\Gamma$. We see that upon increasing temperature the Kondo-Zeno crossover drifts towards lower value of dissipation. In particular we see that the weak dissipation regime is strongly temperature dependent, which is compatible with the interpretation given above that this regime is where Kondo physics remain intact. At large dissipation on the other hand the spin-decay rate is essentially temperature independent. If we now fix the dissipation and increase temperature, see Fig.~\ref{fig:finiteT}(b), we see that the spin decay rate generally increases with temperature, i.e. the dynamics becomes faster due to thermal excitations in the bath. However the dependence from the dissipation rate $\gamma$ reveals a non-monotonic behavior: at low $T$ we see that the spin decay rate increases with $\gamma$, i.e. the decay becomes faster due to the additional decay channels provided by the dephasing. At high temperature on the other hand we see that the decay rate decreases with $\gamma$, as expected from the Zeno effect.  Interestingly, the crossover temperature $T_*$ depends on the interaction strength (See Supplementary Note 3). We can interpret this result  as a finite-temperature signature of the Kondo-Zeno crossover discussed before. Our results imply that the Kondo side of the crossover is strongly dependent on temperature while the Zeno side is robust to finite temperature effects.

{\bf Dot Spectral Function and Kondo Resonance} 

A key quantity to characterize the emergence of the Kondo effect, which is also  experimentally accessible via tunneling spectroscopy, is the dot spectral function $ A_{\sigma  }(\omega)$ defined as the imaginary part of the dot retarded Green's function $G^{R}_{\sigma }(\omega)$~\cite{hewson1993thekondo},
\begin{align}
    A_{\sigma  }(\omega) = -\frac{1}{\pi} \Im \left[ G^{R}_{\sigma }(\omega)\right]\,.
\end{align}
where in the time domain  $G^{R}_{\sigma} (t,t^\prime) = -i \theta (t-t^\prime) \langle \{ d_\sigma(t) , d^\dagger_{\sigma} (t^\prime) \} \rangle$.  We compute the dot Green's function using our NCA dynamical map (See Methods). In Fig.~(\ref{fig:SpectralFuct2})(a-b) we plot the impurity spectral function for $U=4\Gamma$ and increasing values of the dephasing $\gamma$.  First, for $\gamma=0$, we recognize the three-peaks structure of the Anderson impurity spectral function: a narrow Kondo peak at $\omega=0$, signature of a collective entangled many-body state between the dot and the bath which strongly depends on temperature (see Supplementary Note 3), and the incoherent Hubbard bands centered around $\omega=\pm U/2$, describing charge fluctuations. While the NCA results reproduce qualitatively the three-peaks structure of the spectrum, we note that the height of the Kondo peak is suppressed with respect to the exact value.
Adding a small dephasing rate $\gamma<U$ has a strong effect on the Hubbard bands whose position shifts and whose amplitude shrinks down. On the other hand the Kondo peak appears robust to small dephasing, its width and value at zero frequency remaining essentially constant for small $\gamma<U$, see Fig.~\ref{fig:SpectralFuct2}(a). In this regime the spectral function features essentially only a coherent Kondo peak at zero frequency, while charge fluctuations responsible for the Hubbard bands have been quenched by the monitoring process. This is a striking result that is entirely due to the interplay between interaction and dissipation: indeed for a non-interacting dot one can show that dephasing continuously reduces the strength of the coherent peak in the spectral function (see Supplementary Note 3). The robustness of the Kondo peak to weak charge monitoring is consistent with our results on the spin-decay rate $\tau_K^{-1}$ shown in the inset of Fig.~\ref{fig:spin}(c): indeed in the limit  $\gamma\ll \Gamma$ the latter depends weakly on dissipation and so the spectral function (See Fig.~\ref{fig:SpectralFuct2}(a)). 
We emphasize however that a direct connection between spin decay rate and width of the coherent Kondo peak is well established only in absence of dissipation, while for $\gamma\neq0$ other incoherent processes, such as local heating, can be at play in the decay of the spin. Finally, when $\gamma$ is further increased two effects take place in the spectral function (See Fig.~\ref{fig:SpectralFuct2}(b)) namely the value of the zero frequency peak decreases and its width increases, until for $\gamma>U$ the coherent peak at $\omega=0$ disappears and only a broad resonance is left in the spectral function. In the following we provide a physical understanding of the basic mechanism behind the Kondo-Zeno crossover. Before it is useful to discuss the dynamics in the charge sector.
%%%%%%%%%%%%%%%%%%%%%%%%%%%%%%%%%%%%%%%%%%%%%%%%%%%%%%%
\begin{figure*}[!t] 
\centering    
    \includegraphics[width=\textwidth]{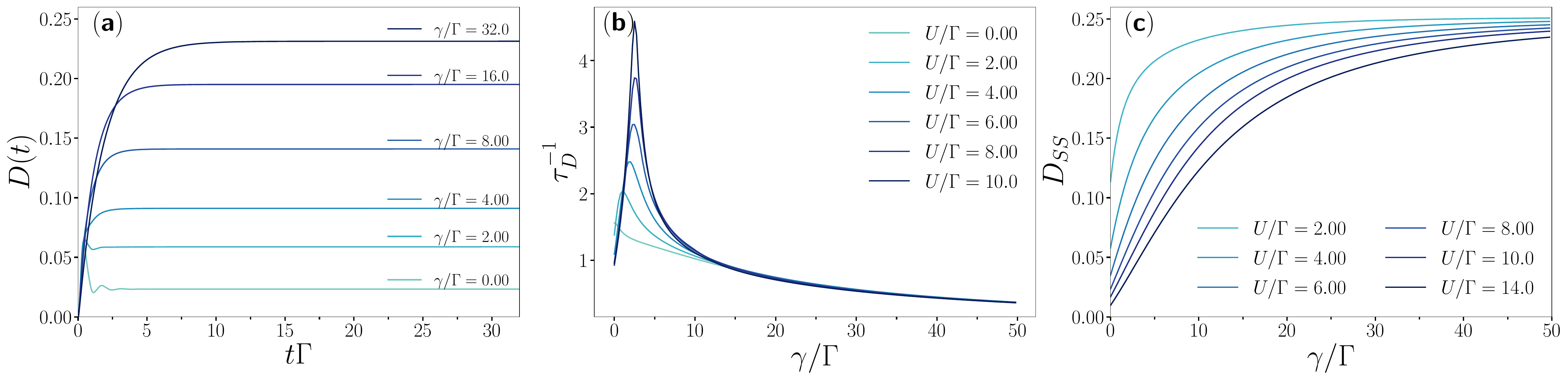}
   \caption{\label{fig:doublons}Dynamics of the doublon fraction starting from a polarised state $\rho_0= \vert \uparrow \rangle \langle  \uparrow \vert $ and evolving under the  dynamics of the half-filled dissipative Anderson Impurity Model. (a) The double occupancy $D(t)$ for $U=8\Gamma$ increases with time and saturates rapidly to a steady-state on a characteristic time scale $\tau_D$. (b) The doublon thermalisation rate $\tau_D^{-1}$ as a function of $\gamma$ shows again signature of the Zeno effect, namely it increases with $\gamma$ through a maximum and then decreases, again as $1/\gamma$. (c) The steady state value of double occupancy $D_{\rm ss}$ increases with $\gamma$, yet we see that for large interaction doublons are suppressed at least for small $\gamma$.}
\end{figure*}
%%%%%%%%%%%%%%%%%%%%%%%%%%%%%%%%%%%%%%%%%%%%%%%%%%%%%%%

{\bf Doublon Dynamics } 

We now consider the dynamics of the dot double occupation $D(t)=\mbox{Tr}\left(\rho_t n_{\uparrow}n_{\downarrow}\right)$, starting again from a singly occupied polarised spin up. In Fig.~\ref{fig:doublons}(a) we plot the dynamics of $D(t)$ for different values of dissipation, at fixed $U=8\Gamma$. While in the unitary case ($\gamma=0$) the doublon fraction displays a non-monotonous time-dependence approaching the steady state value with oscillations, as $\gamma$ increases the dynamics becomes exponential, $D(t)\sim D_{ss}(1-\exp(-t/\tau_D))$ with a characteristic relaxation rate $\tau_D^{-1}$ and a steady-state value $D_{ss}$ that strongly depend on interaction and dephasing. 
In Fig.~\ref{fig:doublons}(b) we see that the doublon relaxation rate also displays a non-monotonic behavior with a sharp maximum and decreases at large $\gamma$ as $1/\gamma$. Comparing Fig.~\ref{fig:doublons}(b) and Fig.~\ref{fig:spin}(c) we see that the doublon relaxation rate is generally larger than the spin one, i.e. doublon dynamics is faster both at zero dissipation (as expected) and for finite $\gamma$. 
The steady-state value $D_{ss}$ increases monotonously with $\gamma$ (see Fig.~\ref{fig:doublons}(c)), and ultimately reaches the value $D_{ss}\sim 1/4$ for $\gamma\rightarrow\infty$,  indicating that locally the impurity is heating up towards infinite temperature. We emphasize however that the steady-state is reached on time scales $t\gg \tau_D$, hence for $\gamma=\infty$ the system remains frozen in the initial spin polarised state, as expected from the continuous Quantum Zeno effect. Strong interactions and coupling to the cold metallic bath however suppress heating by keeping down the doublon fraction, at least for $\gamma < U$.

{\bf Effective Model For Kondo-Zeno Crossover}

In order to understand the emergence of the Kondo-Zeno crossover we derive an effective model that describes our system in the regime when local energy scales, such as interaction $U$ and dissipation $\gamma$ are larger than the coupling to the metallic bath $V_{\bb{k}}$. In equilibrium this is achieved by the standard Schrieffer-Wolff transformation~\cite{hewson1993thekondo} which effectively integrates out charge fluctuations and leads to the Kondo model, describing the antiferromagnetic coupling between the dot's spin and the bath. In the present case the dynamics of the system is non-unitary due to the monitoring process and described by the Lindblad master equation and this has important consequences on the nature of the effective model. In particular, using a dissipative Schrieffer-Wolff transformation (See Supplementary Note 4) on the dissipative Anderson impurity model in Eq~(\ref{eqn:lindblad}) we can obtain an effective non-unitary dynamics at second order in the hybridization. Projecting the dynamics onto the singly occupied sector of the dot, which as we have seen above is justified for strong interactions and weak dissipation, we can show that this non-unitary dynamics is generated by a non-Hermitian Hamiltonian of the form
\begin{align}\label{eq:Kondo}
    H_{\rm eff,K} =- \sum_{\bb{q}\bb{k}} J_{\bb{k}\bb{q}} \left( \Phi_{\bb{q}}^\dagger \frac{\Vec{\sigma}}{2} \Phi_{\bb{k}}\right) \cdot \Vec{S}_{\rm d}
\end{align}
where $\Phi^\dagger_{\bb{k}}=(c_{\bb{k}\uparrow}^\dagger ,c_{\bb{k}\downarrow}^\dagger )$ and $\Vec{S}_{\rm d}=\sum_{\sigma\sigma^{'}}
d^{\dagger}_{\sigma}\Vec{\tau}_{\sigma\sigma^{'}}d_{\sigma^{'}}$ is the dot spin with $\Vec{\tau}_{\sigma\sigma^{'}}$ the Pauli matrices.  Eq.~(\ref{eq:Kondo}) describes a non-Hermitian Kondo model~\cite{Nakagawa2018,hasegawa2021kondo,kattel2024dissipation}, due to the complex-valued spin-exchange coupling $J_{\bb{k}\bb{q}}$ renormalised by dissipation, which reads (by setting $\epsilon_\bb{k} = \epsilon_\bb{q} = 0$ and dropping the momentum dependence of the hybridization $V_{\bb{k}}\simeq V$) 
\begin{align}\label{eqn:JKondo}
    J_{\bb{q}\bb{k}} = J_R+iJ_I=
    %= J = -4 V s_\bb{k} - 4 V x_\bb{k} = 
    - \frac{8 V^2 U}{U^2 + \gamma^2} - i \frac{8 V^2 \gamma}{ U^2 + \gamma^2}
\end{align}
The effective non-Hermitian Hamiltonian captures the competition between Kondo and Zeno screening. This is more clearly seen by considering first the weak dissipation regime $V\ll\gamma\ll U$. Here we have $J_R\gg J_I$ and the system is in the Kondo regime, with few excited doublons that can be safely projected out, yet with a small imaginary Kondo coupling. Results on the non-Hermitian Kondo model suggest a transition from Kondo to a non-Kondo state above a critical value of $J_I$~\cite{Nakagawa2018,lourenco2018kondo,kattel2024dissipation}. This picture is compatible with our NCA results both for the spin-decay rate, which at large $U$ shows robustness against dissipation (See. Fig.~\ref{fig:spin}(b)) and for the impurity spectral function (See. Fig.~\ref{fig:SpectralFuct2}(a)). On the other hand, as we increase the dissipation above $\gamma\sim U$ new physics emerge. First, we see from Eq.~(\ref{eqn:JKondo}) that $J_I$ becomes maximal and then decreases to zero as $1/\gamma$. This is consistent with the spin-decay rate $\tau_K^{-1}$ which therefore in this regime appears to be controlled by dissipation rather than Kondo screening. Furthermore for $\gamma\sim U$ other dissipative channels open up  beyond the single-occupied manifold, such as doublon-holon production terms,  which enter the effective Lindbladian and control the dynamics of the system (See Supplementary Note 4). As confirmed by the NCA results in this regime doublon production becomes substantial and the system locally heats up, leading also to a decay of the initially polarized spin.

Finally we conclude by briefly discussing the role played by charge monitoring. In particular,  we can show that when the monitoring protocol acts on the spin rather than on the charge degrees of freedom of the quantum dot, then the Kondo peak is quickly destroyed by even a weak dissipation (See Supplementary Note 5). 
This result is consistent with our interpretation that the non-trivial Kondo physics in our problem arises because the monitoring protocol does not compete with the spin sector. Interestingly, it can be again understood by means of a Schrieffer-Wolff transformation, where now the effective non-Hermitian Kondo model is supplemented by a dissipative, quantum jump-like,  spin-spin Kondo interaction at the same order as the purely non-Hermitian coupling (See Supplementary Note 5). Such a dissipative coupling, which is also anisotropic thus breaking the  full spin symmetry of the Kondo state,  is likely responsible for the rapid destruction of the Kondo state, even though a full renormalization group study of its interplay with non-Hermitian Kondo couplings is left for future work. This result highlights the key role played by our monitoring protocol to give rise to a non trivial Kondo effect.

\blank
\textbf{Discussion}

In this work we have studied the dissipative dynamics of a strongly correlated quantum dot, described by the Anderson impurity model, under monitoring of the dot's total charge.  We have shown that while charge fluctuations are quenched by the monitoring protocol, the spin dynamics retains a non-trivial many-body dynamics which is strongly affected by the dissipation and displays a crossover in the spin relaxation rate between Kondo screening due to interactions and Zeno effect due to monitoring. This crossover leaves clear signatures at finite temperatures as well as in the steady-state spectral function of the dot, where the Kondo peak at weak monitoring remains robust to dissipation, while Hubbard bands are collapsed due to monitoring of the charge.

The response of a strongly correlated Kondo state to monitoring is a topic that has attracted quite some interest recently. Previous investigations have focused on the effect of projective measurements of singly occupied population on the steady-state spectral function using  weak coupling perturbation theory~\cite{hasegawa2021kondo} or on dissipation due to two-body losses~\cite{stefanini2024dissipative}. Here instead we focus on the dissipative Lindblad dynamics in the strongly correlated regime, treating interaction and dissipation on equal footing via our NCA dynamical map. This allows us to access the full crossover of the spin relaxation rate from Kondo to Zeno regime and to reveal the effect of monitoring on the Kondo peak and the high-energy Hubbard bands in the spectral function. 

We have shown that the Kondo-Zeno crossover can be understood at weak dissipation in terms of a non-Hermitian Kondo model, obtained as effective model at long times through a Schrieffer-Wolff transformation on the Lindbladian. Our interpretation based on the non-Hermitian Kondo model differs significantly from previous attempts that invoke the no-click limit corresponding to full post-selection~\cite{Nakagawa2018,hasegawa2021kondo} to justify this mapping. Here,  it rather emerges from the competition between Coulomb repulsion and charge monitoring: deep in the Kondo regime, when charge fluctuations are frozen due to strong Coulomb repulsion, the leftover local moment can only dissipate through a bath-mediated spin exchange, given in Eq.~(\ref{eqn:JKondo}), while other dissipative processes are blocked at particle-hole symmetry, as long as the doublon production remains limited and the projection onto the single occupied manifold controlled. On the other hand, upon increasing the dephasing rate, heating due to doublon production dominates the spin decay. Our results highlight the robustness of Kondo effect to weak charge monitoring, before heating starts to kick in.  

The Kondo-Zeno crossover can be experimentally observed in quantum dots monitored by a nearby quantum point contact acting as a charge detector~\cite{wiseman2009quantummeasurementand,sankar2024detectortuned}. We note that real-time spin dynamics in quantum dot can be accessed, providing a direct signature of the crossover. In addition, the evolution of the spectral function of the dot under monitoring could be also observed  via tunneling spectroscopy through a second lead~\cite{lebanon2001measuring}. 

This work opens up several avenues for future research. The existence of the Kondo-Zeno crossover scale $\tau_K^{-1}$ raises the question of whether an effective Kondo temperature can be defined for the dissipative Anderson impurity model and if so, what could be the associated scaling theory.  In this respect, a natural future direction, relevant both for its fundamental interest and for experimental applications, would be to look for signatures of Kondo-Zeno crossover in the steady-state transport. Furthermore looking at the dynamics of the conditional state and associated non-linear probes, such as block-entanglement entropy or purity of quantum trajectories, could unveil measurement-induced transitions in this problem.

\blank
\textbf{Methods}\\
\small
\textbf{Hybridization Expansion and Non-Crossing Dynamical Map -  } To tackle the dissipative Anderson Impurity Model we first reformulate the Lindblad many-body dynamics in Eq.~(\ref{eqn:lindblad}) using the purification/superfermion representation~\cite{Prosen_2008,dzhioev2011,HARBOLA2008191,dorda2014auxiliary,Arrigoni2018,werner2023configuration,takahashi96,ojima1981}. This amounts to represent density matrices as pure states $\vert\rho_t\rangle$ in an extended Hilbert space $\mathcal{H}\otimes \tilde{\mathcal{H}}$ that contains a copy of our degrees of freedom. In this formalism the Linbdlad master equation takes the form of a Schrodinger-like equation $\partial_t \vert\rho_t\rangle=\mathcal{L}\vert\rho_t\rangle$ generated by a non-Hermitian operator $\mathcal{L}$, the Lindbladian, which reads~(See Supplementary Note 2)
\begin{align}\label{eqn:Lvector}
    \m{L}= -i \left( H - \Tilde{H}\right) + \gamma \left( n \Tilde{n} - \frac{1}{2} n^2 -\frac{1}{2} \Tilde{n}^2 \right)
\end{align}
where $H$ is given in Eq.~(\ref{eq:H_aim}) and $\tilde{H}$ takes the very same form in terms of fermions living in the $\tilde{\mathcal{H}}$ Hilbert space, $\tilde{c}_{\bb{k},\sigma} ,\tilde{c}^\dagger_{\bb{k},\sigma}$ and $\tilde{d}_{\sigma} ,\tilde{d}^\dagger_{\sigma}$ for the bath and dot operators respectively, which satisfy the usual fermionic algebra (See Supplementary Note 2). We note the analogy between the doubling of the degrees of freedom due to $\tilde{\mathcal{H}}$ and the Keldysh formalism~\cite{secliphd,mcdonald2023third}: the first term in Eq.~(\ref{eqn:Lvector}) can be seen as the coherent evolution along the two independent Keldysh contours while the dissipation term induces  an explicit coupling between them and a back-action (non-Hermitian contribution). We can then perform an exact hybridization expansion in the system-bath coupling, starting from the density matrix in the interaction picture~\cite{Scarlatella2019,scarlatella2023selfconsistent,vanhoecke2024diagrammatic},
%DiagMC samples all  diagrams contributing to
%this bare expansion, which we have shown to enjoy better convergence properties due to dissipation as compared to the unitary case. On the other hand 
and obtain an exact equation for the dressed impurity time-evolution operator $\m{V}(t,0)=\mbox{Tr}_{\rm bath}\left[\exp \left(\m{L}t \right)\right]$, which reads~(See Supplementary Note 2)
\begin{align}
    \partial_t \m{V}(t,0) = \m{L}_{\rm dot} \m{V}(t,0) + \int_0^t d\tau \Sigma(t,\tau) \m{V}(\tau,0)
\end{align}
where $\m{L}_{\rm dot}=-i(H_{\rm dot}-\tilde{H}_{\rm dot})+ \gamma \left( n \Tilde{n} - \frac{1}{2} n^2 -\frac{1}{2} \Tilde{n}^2 \right)$ is the impurity Linbdladian, while the self-energy $\Sigma(t,\tau)$ takes into account the effect of the metallic bath and it is a priori given by an infinite class of one-particle irreducible diagrams. To close the hierarchy we use the NCA dynamical map~\cite{scarlatella2023selfconsistent} corresponding to a self-consistent approximation on the series for $\m{V}$, by keeping only the compact diagrams in which hybridization lines do not cross~\cite{bickersBickers1987b,muller-hartmannMuller-Hartmann1984,meir1993low,nordlander1999how,ecksteinWerner2010a,hartle2013decoherence,erpenbeck2021resolving}
. This amounts to restrict the self-energy to the form $\Sigma(t,\tau)\equiv \Sigma_{NCA}(t,\tau)$ 
%It turns out that the NCA self-energy coincides with the $n=1$ term of the exact self-energy, where the bare propagator $\m{V}_0$ is replaced with the dressed one $\m{V}$. 
%with the NCA self energy given by
\begin{align}\label{eqn:sigmaNCA}
    \Sigma_{NCA}(\tau,\bar{\tau}) &= -i \sum_{\sigma\alpha,\bar{\alpha}}  \Psi^\alpha_\sigma \m{V}\left( \tau,\bar{\tau} \right) \Bar{\Psi}^{\bar{\alpha}}_\sigma \Delta_\sigma^{\alpha \bar{\alpha}}\left( \tau, \bar{\tau}\right) \nonumber\\
    &+i\sum_{\sigma\alpha,\bar{\alpha}} \Bar{\Psi}^{\bar{\alpha}}_\sigma \m{V}\left( \tau,\bar{\tau} \right) \Psi^\alpha_\sigma  \Delta_\sigma^{\alpha \bar{\alpha}}\left( \bar{\tau},\tau\right)
\end{align}
which is itself a function of $\m{V}(t,0)$, hence the self-consistent non-perturbative nature of this method. In Eq.~(\ref{eqn:sigmaNCA}) we have collected the dot operators in $\mathcal{H}$ and $\tilde{\mathcal{H}}$ into a field~(See Supplementary Note 2)
$\bar{\Psi}_{\sigma}=(d^{\dagger}_{\sigma}\; \tilde{d}_{\sigma})$ whose components are labelled by the index $\alpha=0,1$ and introduced the real-time hybridization function $\Delta_{\sigma}^{\alpha\bar{\alpha}}(\tau,\bar{\tau})$, that keeps into account the full non-Markovian nature of the metallic bath (See Supplementary Note 2 for definition.) The resulting dynamical map based on the NCA resummation is then solved self-consistently to compute impurity observables and correlation functions discussed in the manuscript. We conclude by briefly describing how to compute the impurity spectral function within the NCA map. We start from the retarded Green's function of the dot which is defined as
\begin{align}
     G^{R}_{\sigma} (t,t^\prime) = -i \theta (t-t^\prime) \langle \{ d_\sigma(t) , d^\dagger_{\sigma} (t^\prime) \} \rangle
\end{align}
where $\langle \cdots \rangle $ implies the average over the steady state and $\{A,B\} =AB+BA$ is the anti-commutator. The spectral function of the impurity, $A_{\sigma}(\omega)$, is defined by going in Fourier space and taking the imaginary part of the retarded Green's function, i.e. $ A_{\sigma  }(\omega) = -\frac{1}{\pi} \Im \left[ G^{R}_{\sigma }(\omega)\right]$. We can obtain the impurity Green's function from the knowledge of the NCA dynamical map superoperator $\m{V}(t)$ and the impurity steady-state density matrix, a result analog to the quantum regression theorem for Lindblad evolution~\cite{scarlatella2021dynamical}. In particular, using the superfermion representation one can show that
\begin{align}
    G^R_{\sigma } (t,0 ) &=  - i \theta (t) \langle \{ d_\sigma(t), d_{\sigma}^\dagger\} \rangle \notag \\ & = -i \theta (t) \langle I \vert d_\sigma \m{V}(t) d_{\sigma}^\dagger \vert \rho_{\rm dot,ss}  \rangle -\theta(t) \langle I \vert d_\sigma \m{V}(t) \Tilde{d}_{\sigma} \vert \rho_{\rm dot,ss}  \rangle  
\end{align}
In practice we compute first the steady-state density matrix (see Supplementary Note 2) and then solve the NCA equations at long-times to obtain $\m{V}(t)$ and perform Fourier transform to obtain the spectral function.

\textbf{Data availability}\\
The data that support the plots within this paper are provided in the Source Data file. 
%\textbf{Code availability}\\
%All numerical codes in this paper are available upon reasonable request to the authors.

\blank
\textbf{Acknowledgments}\\
This project has received funding from the European Research Council (ERC) under the European Union’s Horizon 2020 research and innovation programme (Grant agreement No. 101002955 — CONQUER). The authors acknowledge computational resources on the Colle\'ge de France IPH cluster.

\clearpage
\appendix
\widetext
\begin{center}
\large{\bf Supplementary Information for `Kondo-Zeno crossover in the dynamics of a monitored quantum dot' \\}
\end{center}
\vspace{2cm}

\setcounter{equation}{0}
\setcounter{figure}{0}
\setcounter{table}{0}
\setcounter{page}{1}
\renewcommand{\theequation}{S\arabic{equation}}
\setcounter{figure}{0}
\renewcommand{\thefigure}{S\arabic{figure}}
\renewcommand{\thepage}{S\arabic{page}}
\renewcommand{\thesection}{S\arabic{section}}
\renewcommand{\thetable}{S\arabic{table}}
\makeatletter

\renewcommand{\thesection}{\arabic{section}}
\renewcommand{\thesubsection}{\thesection.\arabic{subsection}}
\renewcommand{\thesubsubsection}{\thesubsection.\arabic{subsubsection}}

In this Supplemental Information, we discuss: 
\begin{enumerate}
\item The stochastic dynamics of a monitored quantum dot and the Lindblad evolution for the averaged state
\item The hybridization expansion and Non-Crossing Approximation for dissipative quantum impurity models
\item Additional Results: Anderson impurity in the non-dissipative case, spectral function for the dissipative resonant level, impurity entanglement entropy and finite temperature spin relaxation rate.
\item The Schrieffer Wolff transformation and the effective non-Hermitian Kondo model
\item The dynamics of the Anderson Impurity Model under spin dephasing and its effective Kondo model
\end{enumerate}

\section{Supplementary Note 1: Lindblad Dynamics from continuous monitoring}

We consider a quantum dot coupled to a metallic lead and to a monitoring device which measures some Hermitian operator $O$. We focus here on a homodyne detection scheme corresponding to a quantum-state diffusion protocol described by the following stochastic Schrodinger equation~\cite{wiseman2009quantummeasurementand}
\begin{align}\label{eq:qsd}
        d\vert\psi(\xi_t)\rangle &= -idt H\vert\psi(\xi_t)\rangle+d\xi_t\left(O-\langle O\rangle\right)\vert\psi(\xi_t)\rangle-\frac{\gamma dt}{2}\left(O-\langle O\rangle\right)^2\vert\psi(\xi_t)\rangle
\end{align}
Here we follow the standard terminology and define the state $\vert\psi(\xi_t)\rangle$ as the ensemble of quantum trajectories, each labeled by a specific history of measurement outcomes $\xi_t$.
The first term in Eq.~(\ref{eq:qsd}) describes the deterministic evolution, a coherent dynamics generated by the dot+lead Hamiltonian
\begin{align}
H=\sum_{\bb{k},\sigma}\varepsilon_{\bb{k}}c^\dagger_{\bb{k},\sigma}c_{\bb{k},\sigma}+ \sum_{\bb{k},\sigma} \left(V_{\bb{k}} d_{\sigma}^\dagger c_{\bb{k},\sigma} + h.c \right)+
\varepsilon_d \sum_{\sigma}d^\dagger_{\sigma}d_{\sigma}+Un_{\uparrow}n_{\downarrow}
\end{align}
The continuous monitoring of the operator $O$ has two effects on the evolution of the state of the system, as seen in the last terms in 
Eq.~(\ref{eq:qsd}). It introduces a local stochastic term whose noise couples to the fluctuations $O-\langle O\rangle$, as well as a back-action term proportional to the fluctuation squared, $(O-\langle O\rangle)^2$. Here, we defined the expectation value $\langle \circ\rangle_\xi =\langle \psi(\xi_t)\vert \circ\vert\psi(\xi_t)\rangle$ and the stochastic real variable $d\xi_t$, satisfying the standard rules from \^Ito calculus $\overline{d\xi_t} = 0 $ and $(d\xi_t)^2  =\gamma dt $.  The stochastic variable $d\xi_t$ can be interpreted as an infinitesimal fluctuating noise term. 

From the stochastic Schrodinger equation in Eq.~(\ref{eq:qsd}) one can obtain the density matrix $\rho(\xi_t)=\vert \psi(\xi_t)\rangle\langle \psi(\xi_t)\vert$ and write down an equation of motion for its evolution. Using the properties of the  \^Ito noise this reads
\begin{align}\label{eqn:drho}
    d\rho(\xi_t)=-i[H,\rho(\xi_t)]+\gamma O\rho(\xi_t) O-\frac{\gamma }{2}\left\{\rho(\xi_t),O^2\right\}+
    d\xi_t\left\{O-\langle O\rangle_{\xi},\rho(\xi_t)\right\}
\end{align}

Due to the properties of the  \^Ito noise which is uncorrelated at different times, the average of Eq.~(\ref{eqn:drho}) yields the following Lindblad equation for the averaged state $\rho_t\equiv \overline{\rho(\xi_t)}$
\begin{equation}
   \partial_t \rho_t = -i \left[ H, \rho_t \right] +  \gamma O\rho_t O -\frac{\gamma }{2} \{ O^2 , \rho_t \}
\end{equation}
which coincides with the one given in the main text for Hermitian jump operators.

\section{Supplementary Note 2: Hybridization Expansion and Non-Crossing Diagrams}

\subsection{Vectorization of the Lindbladian}

We start by applying the superfermion or vectorization formalism~\cite{dzhioev2011,dorda2014auxiliary} to the case of the master equation for a dissipative quantum impurity model, i.e.
\begin{align}
    \partial_t \rho_t = -i \left[ H, \rho_t \right] + \gamma L \rho_t L^\dagger -\frac{\gamma}{2} \{ L^\dagger L , \rho_t \}
\end{align}
where $H=H_{\rm dot}+H_{\rm bath}+H_{\rm hyb}$ is the dot plus bath Hamiltonian given in the main text while $L$ is the impurity jump operator, corresponding in our case to the dot total charge $L=\sum_{\sigma}n_{\sigma}$. We introduce the Hilbert spaces $\m{H}$ for the impurity and the bath degrees of freedom and its doubled tilde-version $\tilde{\m{H}}$ and the associated identity operators
\begin{align}
    I  &= \sum_{m} \vert m \rangle  \langle m \vert\\
    \tilde{I} &=\sum_{m} \vert \tilde{m} \rangle  \langle \tilde{m} \vert \,.
\end{align}
written in terms of two orthonormal basis $\vert m\rangle,\vert \tilde{m} \rangle$. We can also duplicate all the degrees of freedom in the problem, namely the impurity and the bath fermions, and introduce the associated creation/annihilation operators $d_{\sigma},\tilde{d}_\sigma$ and $c_{\bb{k},\sigma},\tilde{c}_{\bb{k},\sigma}$ and their Hermitian conjugate. The key step is now to vectorize the identity operator, introducing the so called left vacuum~\cite{dorda2014auxiliary,dzhioev2011} (or vectorized identity)
\begin{align}
    \vert I  \rangle = \sum_{m} \left( -i \right)^{m} \vert m \rangle \otimes | \tilde{m} \rangle
\end{align}
The vectorized identity is particularly useful as it allows to write any operator $O$ as a vector
\begin{align}
    \vert O \rangle = O \vert I \rangle = O \otimes \Tilde{I} \vert I\rangle
\end{align}
In particular, the vectorized density matrix reads:
\begin{align}
    \vert \rho \rangle = \rho \vert I \rangle
\end{align}
Using the left vacuum one can write down the trace of any operator over the density matrix as 
\begin{align}
\langle O(t)\rangle=\mbox{Tr}\left(\rho_t O\right)\equiv 
\langle I\vert O\vert\rho_t\rangle
\end{align}
Following these rules we can rewrite the Linblad master equation as a non-unitary Schrodinger type of equation
\begin{align}\label{eqn:partial_trho}
\partial_t\vert\rho_t\rangle=\mathcal{L}\vert\rho_t\rangle
\end{align}
where  $\mathcal{L}$ is the vectorized Lindbladian. The advantage of the vectorization formalism is that the superoperator structure usually needed to treat Lindbladian problems and the associated hybridization expansion is now encoded by doubling the local Hilbert space and working with an additional quantum number, similar to an orbital degrees of freedom.
The Lindbladian $\mathcal{L}$ has now three contributions
\begin{equation}\label{eqn:L_vect}
    \m{L} = \m{L}_{\rm dot} + \m{L}_{\rm bath}+\m{L}_{\rm hyb}
\end{equation}
The first one $\m{L}_{\rm dot}$ is the free Lindbladian for the dissipative impurity. By using the super-fermions rules~\cite{ojima1981,dorda2014auxiliary} ($d_\sigma | I \rangle  = -i \tilde{d}^\dagger_\sigma |I\rangle$ and $d_\sigma^\dagger | I \rangle  = -i \tilde{d}_\sigma |I\rangle$) and since we consider only the dissipation on the impurity degrees of freedom, we can formally write the impurity  Lindbladian $\m{L}_{\rm dot}$ as:
%\begin{widetext}
    \begin{equation}
    \m{L}_{\rm dot} = -i\left(H_{\rm dot} - \Tilde{H}_{\rm dot}  \right) + 
    \gamma \left(s_L L \Tilde{L} - \frac{1}{2} L^\dagger L - \frac{1}{2} \Tilde{L}^\dagger \Tilde{L} \right)
    %\sum_\mu \left(s_{L_\mu} L_\mu \Tilde{L}_\mu - \frac{1}{2} L_\mu^\dagger L_\mu - \frac{1}{2} \Tilde{L}_\mu^\dagger \Tilde{L}_\mu \right)
\end{equation}
%\end{widetext}
where $s_{L}$ is an extra sign depending on the fermionic ($s_{L}=-i$) or bosonic ($s_{L}=1$) nature of the jumps operator. The second term in Eq.~(\ref{eqn:L_vect}) 
is the bath Lindbladian and reads $\m{L}_{\rm bath}=-i\left(H_{\rm bath}-\tilde{H}_{\rm bath}\right)$. Finally 
the impurity-bath Lindbladian $\m{L}_{\rm hyb}$ can be written in compact form by introducing the following fields
\begin{align}\label{eqn:phi_psi}
    \Phi_{\sigma} = \sum_{\bb{k}} V_\bb{k} \begin{pmatrix}
        c_{\bb{k},\sigma} \\ 
        \Tilde{c}_{\bb{k},\sigma}^\dagger
    \end{pmatrix} \quad \Psi_\sigma = \begin{pmatrix}
        d_{\sigma} \\ 
        \Tilde{d}_{\sigma}^\dagger
    \end{pmatrix}
\end{align}
which group together the operators living in the space $\m{H}$ and $\tilde{\m{H}}$. Using these fields we can write the system-bath term in a more compact way:
\begin{equation}\label{eqn:L_SB}
    \m{L}_{\rm hyb} =-i\left(H_{\rm hyb}-\tilde{H}_{\rm hyb}\right) =  -i \sum_{\sigma\alpha} \left( \Bar{\Phi}^{\alpha}_{\sigma} \Psi^{\alpha}_\sigma + \Bar{\Psi}^{\alpha}_\sigma \Phi^{\alpha}_{\sigma} \right)
\end{equation}
where we have introduced a label $\alpha=0,1$ which denotes the Hilbert space $\m{H}$ or $\tilde{\m{H}}$ ($d_\sigma = \Psi_\sigma^{\alpha=0}$ and $\tilde{d}^\dagger_\sigma = \Psi_\sigma^{\alpha=1}$). At this point we can write the formal solution of the vectorized master equation~(\ref{eqn:partial_trho}) 
\begin{equation}\label{eqn:rho_t}
    \vert \rho_t \rangle = \m{T}_t \exp\left(  \int_0^t \m{L}(\tau) d\tau \right) \vert \rho_0 \rangle 
\end{equation}
where we have introduced the time ordering operator $\m{T}_t$ in the Superfermions representation. Unlike the standard Keldysh contour, here the time ordering is defined as:
\begin{align}
    t_{\alpha} > \bar{t}_{\beta} = \left\{ \begin{array}{ll}
    t>\bar{t} \quad \text{when} \quad \alpha = \beta \in \m{H},\Tilde{\m{H}} \\ 
   \forall t,\bar{t} \quad \text{when}\quad  \alpha \in \m{H} \quad \beta \in \tilde{\m{H}}
    \end{array} \right.
\end{align}
namely it corresponds to the regular time-ordering if $\alpha=\beta$, while if these indexes are different and $\alpha$ belongs to $\m{H}$ then $t_{\alpha}$ is always greater than $\bar{t}_{\beta}$. In the opposite case, $\alpha$ belonging to $\tilde{\m{H}}$ would imply that $ t_{\alpha}$ is never greater than $\bar{t}_{\beta}$. This ordering allows to define a time-ordering operator $\m{T}_t$ such that two operators, $\psi_1$ and $\psi_2$, being $\psi$ a creation or annihilation fermionic operator living in the $\m{H}(\Tilde{\m{H}})$ Hilbert space, anticommute under time-ordering:
\begin{align}
    \m{T}_t \psi_1(t_\alpha) \psi_2(t_\beta) = \left \{ \begin{array}{ll}
    \psi_1(t_\alpha) \psi_2(t_\beta) \quad \text{if} \quad t_\alpha > t_\beta \\
    - \psi_2(t_\beta) \psi_1(t_\alpha) \quad \text{otherwise 
} \end{array} \right.
\end{align}
Eq.~(\ref{eqn:rho_t}) represents the starting point to perform the hybridization expansion, namely an expansion order by order in the system-bath coupling $\mathcal{L}_{\rm hyb}$, as we will discuss in the next section.
%We now want to perform a partial trace over the bath degrees of freedom to define a reduced dynamics for the system, generated by a reduced (system only) evolution operator. 

%In order to obtain the partial trace over the bath degrees of freedom in the vectorized formalism we define the modified scalar product:
%\begin{equation}
%    \langle I_B | \rho(t) \rangle = \langle I_B | \m{V}(t,0) | \rho_0 \rangle
%\end{equation}
%where we have introduced the left-vacuum restricted to the bath degrees of freedom only, i.e
%\begin{equation}
%    |I_B \rangle = \sum_{a \in \m{B}} | a\rangle | \Tilde{a} \rangle
%\end{equation}
%Under this assumption and considering a factorized initial density matrix $\rho_0 = \rho_{0,I} \otimes \rho_{0,B}$ we obtain the reduced evolution operator:
%\begin{equation}
%    \m{V}_s(t,0) = \langle I_B | \m{V}(t,0) | \rho_{0,B} \rangle
%\end{equation}
%which clearly lives only in the impurity space as expected. note that $| \rho_{0,B} \rangle$ is the initial density matrix of the bath. And in the last expression we have assumed a factorized initial density matrix $\rho_0 = \rho_{0,B} \otimes \rho_{0,s}$.

\subsection{Hybridization Expansion and NCA Dynamical Map}

In this Section we derive the perturbation series (called Hybridization expansion) for the evolution superoperator in powers of the impurity-bath coupling, up to all orders in this coupling. In the interaction picture we can write the time evolution operator as
\begin{align}
    | \rho_t \rangle = \m{V}_0(t) \m{T}_t exp \left( \int_0^t \m{L}_{\rm hyb}(s)ds\right) | \rho_0 \rangle 
\end{align}
where $\m{V}_0(t)=\exp\left( \m{L}_0 t\right)$ is the  time evolution operator in the decoupled limit, i.e. $\m{L}_0=\m{L}_{\rm dot}+\m{L}_{\rm bath}$, while the time dependence of $\m{L}_{\rm hyb}$ is given  by the free evolution $\m{L}_{\rm hyb}\left( \tau \right) = e^{-\m{L}_0 \tau} \m{L}_{\rm hyb} e^{\m{L}_0 \tau}$. By using a standard Taylor expansion for the bath-impurity terms we can write all the order of $|\rho_t \rangle$ 
\begin{align}
    |\rho_t \rangle = \sum_n \frac{1 }{n!} \int_0^t d\tau_1 \cdots d\tau_n \m{V}_0(t) \m{T}_t \left[ \m{L}_{\rm hyb}(\tau_1) \m{L}_{\rm hyb}(\tau_2) \cdots \m{L}_{\rm hyb}(\tau_n)  \right] |\rho_0\rangle
\end{align}
Due to the fact that the coupling between the bath and the impurity degrees of freedom is linear, we only have to consider the even terms of the expansion
\begin{align}
    |\rho_t \rangle  = \sum_n \frac{1 }{2n!} \int_0^t d\tau_1 \cdots d\tau_{2n} \m{V}_0 (t) \m{T}_t \left[ \m{L}_{\rm hyb}(\tau_1) \m{L}_{\rm hyb}(\tau_2) \cdots \m{L}_{\rm hyb}(\tau_{2n})  \right] |\rho_0\rangle\,.
\end{align}
Finally, in terms of the spinors $\Psi$ and $\Phi$, the time evolution of the density matrix becomes 
\begin{align}
    |\rho_t \rangle  = \sum_n \sum_{\{\sigma ,\bar{\sigma }\}}\frac{(-1)^n }{(n!)^2} \int_0^t d\tau_1 \cdots d\tau_{n} \int_0^t d\bar{\tau}_1 \cdots d\bar{\tau}_{n} \m{V}_0 \m{T}_t \left[ \bar{\Psi}_\sigma(\tau_1)\Phi_\sigma(\tau_1) \cdots \bar{\Phi}_\sigma(\tau_n)\Psi_\sigma(\tau_n)  \right] |\rho_0\rangle
\end{align}
We now take the average over the bath $Tr_{\rm bath}\left[\cdots\right]= \langle I_{\rm bath} | \cdots | \rho_0 \rangle$ and use the fact that the initial state
$|\rho_0 \rangle$ is factorized, i.e. initially we take the initial density matrix as $\rho_0 = \rho_{\rm dot,0} \otimes \rho_{\rm bath,0} $. Then, since the bath is non-interacting, we can use the Wick theorem in order to trace out the bath degrees of freedom and obtain the hybridization expansion for the dressed impurity propagator $\langle I_{\rm bath}|\rho_t \rangle \equiv \m{V}(t,0)\vert\rho_{\rm dot,0}\rangle$ which finally reads~\cite{vanhoecke2024diagrammatic}
\begin{align}\label{eqn:hyb_exp}
   \m{V}(t,0)\vert\rho_{\rm dot,0}\rangle =\sum_n \sum_{\{\sigma ,\bar{\sigma }\}}\frac{(-i)^n }{(n!)^2} \int_0^t \prod_i d\tau_i d\bar{\tau}_i
   \exp\left( \m{L}_{\rm dot} t\right)
   \m{T}_t \left[ \bar{\Psi}_\sigma(\tau_1) \cdots \Psi_\sigma(\tau_n)  \right] |\rho_{\rm dot,0}\rangle \rm Det_\sigma \left[\{\bar{\Phi}_\sigma ,\Phi_\sigma \} \right] \rm Det_{\bar{\sigma}} \left[\{\bar{\Phi}_{\bar{\sigma}} ,\Phi_{\bar{\sigma}} \} \right] 
\end{align}
where $\rm Det_\sigma \left[\{\bar{\Phi}_{\bar{\sigma}} ,\Phi_{\bar{\sigma}} \} \right] $ are determinants built out of the real-time hybridization functions of the bath defined as
\begin{align}
    \Delta_\sigma^{\alpha \bar{\alpha}}(\tau ,\bar{\tau}) = -i \langle I_{\rm bath} |  \m{T}_t \left[ \bar{\Phi}_\sigma^\alpha (\tau) \Phi_\sigma^{\bar{\alpha}}(\bar{\tau}) \right]|\rho_{\rm bath,0} \rangle
\end{align}
For a non-interacting bath, as the one we consider here, a simple calculation gives~\cite{vanhoecke2024diagrammatic} 
\begin{align}
    \Delta_\sigma^{01}(\tau,\bar{\tau}) &= \int d\epsilon \left(1-n_F(\epsilon) \right)\Gamma_\sigma(\epsilon) e^{- i \epsilon (\tau -\bar{\tau})}\\
     \Delta_\sigma^{10}(\tau,\bar{\tau}) &= -\int d\epsilon n_F(\epsilon) \Gamma_\sigma(\epsilon) e^{- i \epsilon (\tau -\bar{\tau})}
\end{align}
where $n_F(\epsilon)$ is the Fermi distribution and $\Gamma(\epsilon)$ the energy-dependent hybridization for the spin channel $\sigma$, defined in the main text. From this result the diagonal components of $\Delta_\sigma^{\alpha \bar{\alpha}}(\tau ,\bar{\tau})$ can be obtained, for example $\Delta_\sigma^{0}(\tau,\bar{\tau})=i \theta(\tau -\bar{\tau}) \Delta_\sigma^{10}(\tau,\bar{\tau}) + i \theta( \bar{\tau}-\tau ) \Delta_\sigma^{01}(\tau,\bar{\tau})
$.

The hybridization expansion we have derived in Eq.~(\ref{eqn:hyb_exp}) is a bare expansion in the dot-bath coupling. It can be reorganized in a self-consistent diagrammatic expansion by identifying one-particle irreducible diagrams and introducing a self-energy $\Sigma(\tau,\tau^\prime)$ to obtain a Dyson equation for $\m{V}(t,0)$ of the form~\cite{scarlatella2023selfconsistent}\begin{align}\label{eqn:dynamicalmap}
    \m{V}(t,0) = \m{V}_{\rm dot}(t,0) + \int_0^t d\tau \int_0^\tau d \bar{\tau} \m{V}_{\rm dot} \left( t, \tau \right) \Sigma \left( \tau,\bar{\tau} \right) \m{V} \left( \bar{\tau}, 0\right)
\end{align}
which can be rewritten as in the main text upon taking a time-derivative on both sides of the equation and using $\partial_t \m{V}_{\rm dot}(t,0)=\m{L}_{\rm dot}\m{V}_{\rm dot}(t,0)$. As in any interacting diagrammatic theory the self-energy is not known in closed form. It can however reorganised as an expansion in diagrams with increasing number of crossing hybridization lines~\cite{scarlatella2021dynamical,scarlatella2023selfconsistent}. The lowest order self-consistent level of this hierarchy is the so called non-crossing approximation (NCA) which corresponds to keeping only the compact diagrams in which hybridization lines do not cross. The formal expression for the NCA self-energy is therefore the one given in the main text, which reads
%we can replace the expression of the Impurity-Bath lindbladian $\m{L}_{SB}$,
%\begin{align}
%    |\rho_t \rangle  = \sum_n \sum_{\{\sigma ,\bar{\sigma }\}}\frac{(-1)^n }{(n!)^2} \int_0^t d\tau_1 \cdots d\tau_{n} \int_0^t d\bar{\tau}_1 \cdots d\bar{\tau}_{n} \m{V}_0 \m{T}_t \left[ \bar{\Psi}_\sigma(\tau_1)\Phi_\sigma(\tau_1) \cdots \bar{\Phi}_\sigma(\tau_n)\Psi_\sigma(\tau_n)  \right] |\rho_0\rangle
%\end{align}
\begin{align}\label{eq:sigmaNCA_sup}
    \Sigma_{NCA}(\tau,\bar{\tau}) = -i \sum_\sigma \sum_{\alpha,\bar{\alpha}} \left[ \Psi^\alpha_\sigma \m{V}\left( \tau,\bar{\tau} \right) \Bar{\Psi}^{\bar{\alpha}}_\sigma \Delta_\sigma^{\alpha \bar{\alpha}}\left( \tau, \bar{\tau}\right) - \Bar{\Psi}^{\bar{\alpha}}_\sigma \m{V}\left( \tau,\bar{\tau} \right) \Psi^\alpha_\sigma  \Delta_\sigma^{\alpha \bar{\alpha}}\left( \bar{\tau},\tau\right) \right]
\end{align}
which describes a rainbow diagram with a dot vertex 
$\bar{\Psi}_{\sigma}^{\bar{\alpha}}$ at time $\bar{\tau}$, a dressed dot propagation from $\bar{\tau}$ to $\tau$ with propagator $\m{V}(\tau,\bar{\tau})$ and a hybridization line $\Delta_\sigma^{\alpha \bar{\alpha}}\left( \tau, \bar{\tau}\right)$ and finally another dot vertex operator $\Psi_{\sigma}^{\alpha}$ at time $\tau$, with the second term in Eq.~(\ref{eq:sigmaNCA_sup}) describing the diagram with $\tau,\bar{\tau}$ exchanged.

\subsubsection{Steady-state Condition}
%\textcolor{red}{check this section}
\begin{figure*}[!t] 
 \includegraphics[width=0.9\textwidth]{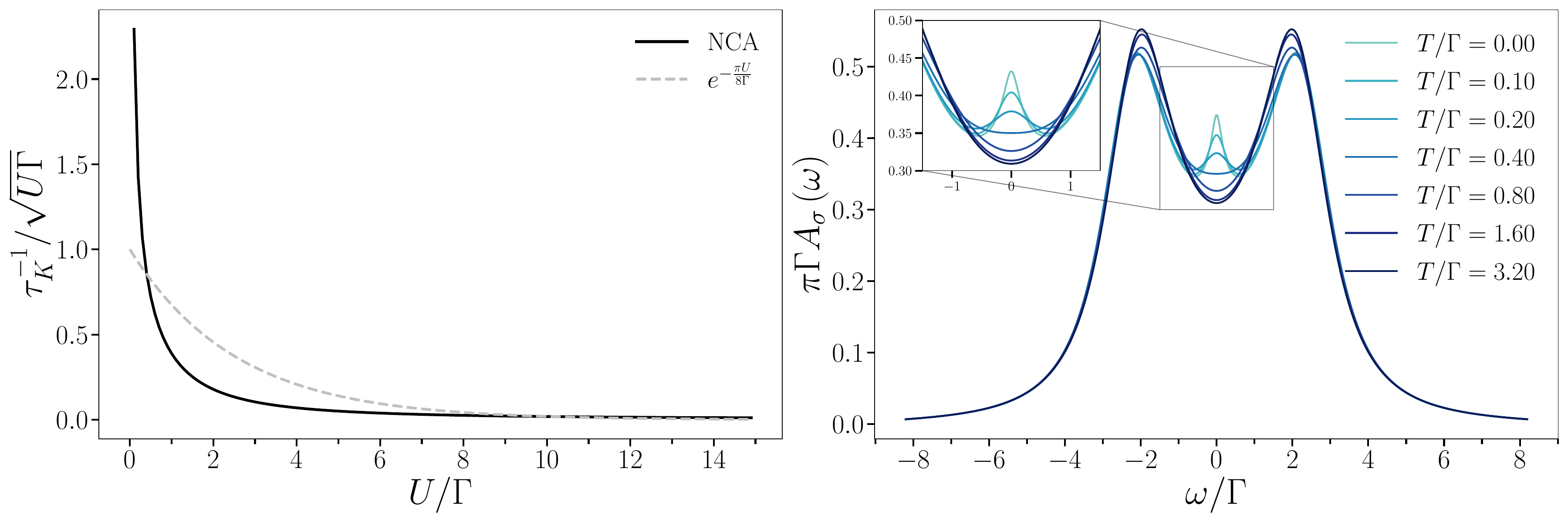}
    \caption{\label{fig:benchmarkNCA} NCA results for the non-dissipative Anderson Impurity Model, corresponding to $\gamma=0$. (a) Spin relaxation rate as a function of interaction $U$. (b) Impurity spectral function for the half-filled case at fixed $U=5\Gamma$ and different temperatures.}
\end{figure*}

Within NCA we can obtain an equation directly for the steady-state density matrix of the impurity~\cite{scarlatella2023selfconsistent}, which is useful for example when computing spectral functions. If we assume a unique steady state, the associated density matrix is defined as $| \rho_{\rm dot,ss} \rangle = \m{V}(\m{1},0) | \rho_{\rm dot,0} \rangle$. While in principle this would require to perform the full transient dynamics from an arbitrary initial condition, here we show how to obtain $\vert \rho_{ss}\rangle$ directly from the stationary state propagator $\m{V}\left(t,t^\prime \right)$.
\\
\\
By definition, the steady-state density matrix satisfies the condition
\begin{align}
    \partial_t \m{V}(t,0) |\rho_{\rm dot,0}\rangle \underset{t\rightarrow \m{1}}{\longrightarrow} 0
\end{align}
where $\rho_0$ is an arbitrary initial state. By using the Dyson equation we get:
\begin{align}
    \underset{t \rightarrow \m{1}}{lim} \partial_t \m{V}(t,0) |\rho_{\rm dot,0}\rangle  = \underset{t \rightarrow \m{1}}{lim} \left( \m{L}_{\rm dot} \m{V}(t,0) + \int_0^t d\tau \Sigma_{NCA}(t,\tau) \m{V}\left( \tau,0 \right)\right) |\rho_{\rm dot,0} \rangle =0
\end{align}
By making a change of variable in the convolution integral the condition reads
\begin{align}
    \underset{t \rightarrow \m{1}}{lim} \left( \m{L}_{\rm dot} \m{V}(t,0) + \int_0^t d\tau \Sigma_{NCA}(t,t-\tau) \m{V}\left( t-\tau,0 \right)\right) |\rho_{\rm dot,0} \rangle =0
\end{align}
We have now to assume some hypothesis, the first one is that the system is supposed to loose memory of the initial conditions, this mean that the self energy must vanish when $\tau \approx t \rightarrow \m{1}$. Under this hypothesis the convolution integral in the Dyson equation can be cutoff when $t-\tau > t_{memory}$. Then when the self energy is non-zero the propagator $\m{V}\left( t-\tau,0 \right)$ is stationary and can be replaced by $\m{V}(\m{1},0)$. So under the previous hypothesis, the steady state density matrix satisfies:
\begin{align}\label{eqn:rho_ss}
    \left(\m{L}_{\rm dot} + \int_0^\m{1} \Sigma_{NCA}\left( \tau\right) d\tau \right)|\rho_{\rm dot,ss} \rangle =0
\end{align}
This condition depends only on the stationary state propagator $\m{V}(\tau)$ through the NCA self-energy $\Sigma_{NCA}(\tau)$.

\section{Supplementary Note 3: Additional Results}

In this Section we present additional NCA results on the dissipative Anderson Impurity Model: (i) benchmark of the non-dissipative case; (ii) benchmark of the impurity spectral function in the dissipative but non-interacting case (dissipative resonant level model); (iii) the impurity entanglement entropy ; (iv) the spin relaxation rate at finite temperature.

\subsection{NCA Results for the non-dissipative Anderson Impurity Model}

We start presenting  NCA results for the non-dissipative Anderson Impurity Model, corresponding to the limit $\gamma=0$ of the model discussed in this work. This model can be still solved with our NCA dynamical map, the main difference is that the local impurity problem is now described by Hamiltonian evolution rather than Lindblad. We consider first the spin relaxation dynamics, namely we start from a factorized initial density matrix where the impurity is polarized and the bath is in equilibrium at zero temperature and suddenly switch-on the hybridization, leading to the dynamics of the Kondo effect. We compute the evolution of the dot magnetization as a function of time which relaxes exponentially to zero and extract the spin-relaxation rate or inverse Kondo time, $\tau_K^{-1}$ that we plot in Fig.~(\ref{fig:benchmarkNCA})(left) as a function of interaction $U$. We show that the relaxation rate is strongly suppressed at large interactions, with a behavior that well reproduces the exponential scaling to zero of the Kondo temperature (see dashed line). In the right panel of Fig.~(\ref{fig:benchmarkNCA} we compute the impurity spectral function for different temperatures and demonstrate the destruction of the Kondo peak due to thermal effects. We note that the size of the coherent Kondo peak is quantitatively underestimated in NCA. Going beyond the NCA approximation in our self-consistent dynamical map, i.e. including the one-crossing diagrams~\cite{scarlatella2023selfconsistent}, would improve this issue and it is a priori possible, although computationally more involved in the dissipative fermionic case.

\begin{figure*}[!t] 
 \includegraphics[width=1.0\textwidth]{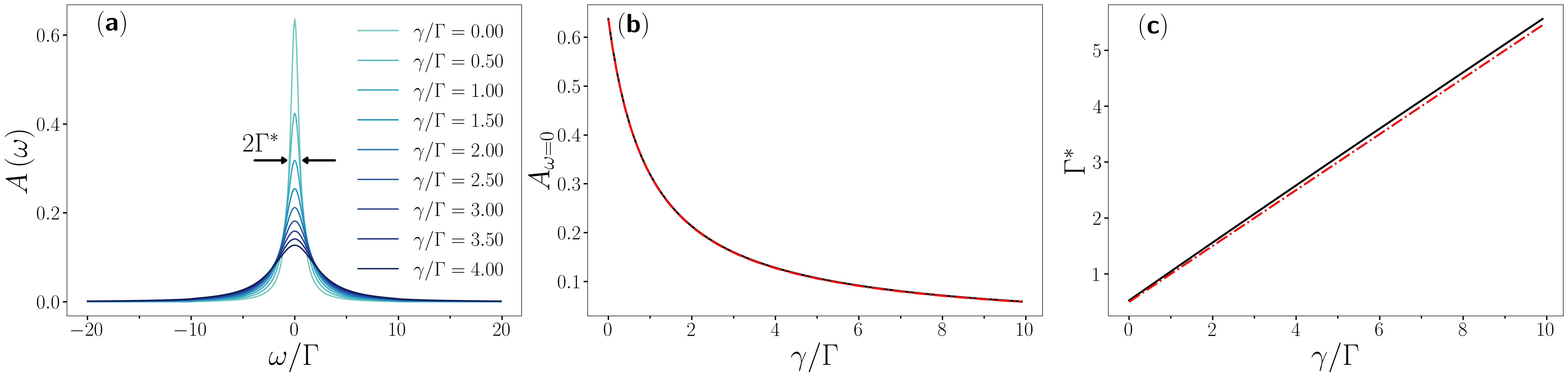}
    \caption{\label{fig:SpectralFuct} Dissipative Resonant Level Model - (a) Impurity spectral function for the half-filled dissipative resonant level model, corresponding to $U = -2 \epsilon_d = 0 \Gamma $ and increasing values of dephasing $\gamma$. (b) Spectral function value at zero frequency as a function of $\gamma$. (c) Width $\Gamma^*$ of the zero frequency peak as a function of $\gamma$. In both cases the agreement between NCA (solid line, for $W=10 \Gamma$) and the exact solution (dashed line, see main text) is excellent.}
%    Spectral function of the impurity in the Half filling case $U = -2 \epsilon_d = 0 \Gamma $. The red dot line corespond to the analytical result and the black solid line the NCA dynamics in the limit $W=10 \Gamma$}
\end{figure*}

\subsection{Impurity Spectral Function for the Dissipative Resonant Level Model}

We now consider the dissipative but non-interacting case, $U=0$, corresponding to a dissipative Resonant Level model. We stress that due to the monitoring of the total charge the model is still interacting, i.e. non-gaussian. However the specific nature of the dissipative interaction makes it possible to close exactly the equations of motion for the retarded Green's function in terms of a Dyson equation. We obtain therefore
\begin{align}
    G^R_\sigma(\omega) = \frac{1}{(G^R_{0\sigma})^{-1} - \Sigma^R_\sigma(\omega)}
\end{align}
where the bare retarded Green's function of the impurity reads $(G^R_{0\sigma})^{-1}=\omega - \epsilon_d + i \eta $, while the self-energy $\Sigma^R_{\sigma} (\omega)$
\begin{align}
    \Sigma^R_{\sigma} (\omega)  = -i \frac{\gamma}{2} +\sum_{\bb{k}} \frac{\vert V_\bb{k} \vert^2}{\omega - \epsilon_\bb{k} + i\eta } = \m{PP} \sum_\bb{k} \frac{\vert V_\bb{k} \vert^2}{\omega-\epsilon_\bb{k} } - \frac{i}{2} \left[ \gamma + \Gamma(\omega) \right]
\end{align}
where $\Gamma(\epsilon) = 2\pi \sum_\bb{k} \vert V_\bb{k} \vert^2\delta(\epsilon - \epsilon_\bb{k})$ is the hybridization function.
One can therefore obtain the spectral function in closed form in the wide-bandwidth limit
\begin{align}\label{eqn:exact_aomega}
    A_\sigma (\omega) = - \frac{1}{\pi} Im \left[ G^R_\sigma (\omega)\right]   = \frac{1}{2\pi}\frac{(\gamma+ \Gamma)}{(\omega - \epsilon_d)^2 + (\gamma+ \Gamma)^2 / 4 }
\end{align}
From this expression we can immediately read out the renormalised width $\Gamma^*$,
\begin{align}
    \Gamma^\ast = \frac{1}{2} \left(\gamma + \Gamma \right)
\end{align}
In Fig.~(\ref{fig:SpectralFuct})(a) we plot the spectral function obtained with NCA in the $U=0$ case for increasing value of the dephasing rate. We see that the resonance at the Fermi level is broadened and at the same time the value at $\omega=0$ decreases with $\gamma$. A comparison with the exact result for the $U=0$ case is shown in Fig.~(\ref{fig:SpectralFuct})(b-c),  concerning the weight at $\omega=0$ and the width of the resonance, demonstrating the perfect agreement between NCA and the exact result. We note that from the exact expression in Eq.~(\ref{eqn:exact_aomega}) we conclude that the value of the spectral function at $\omega=0$ is affected by the dissipative interaction (monitoring rate/dephasing), even though the system remains half-filled. This is different than in the unitary case, where for $U\neq0$ the value of the zero frequency spectral function is not renormalised and points towards a breakdown of the Luttinger theorem for dissipative quantum impurity models.

\begin{figure*}[!t] 
 \includegraphics[width=0.95\textwidth]{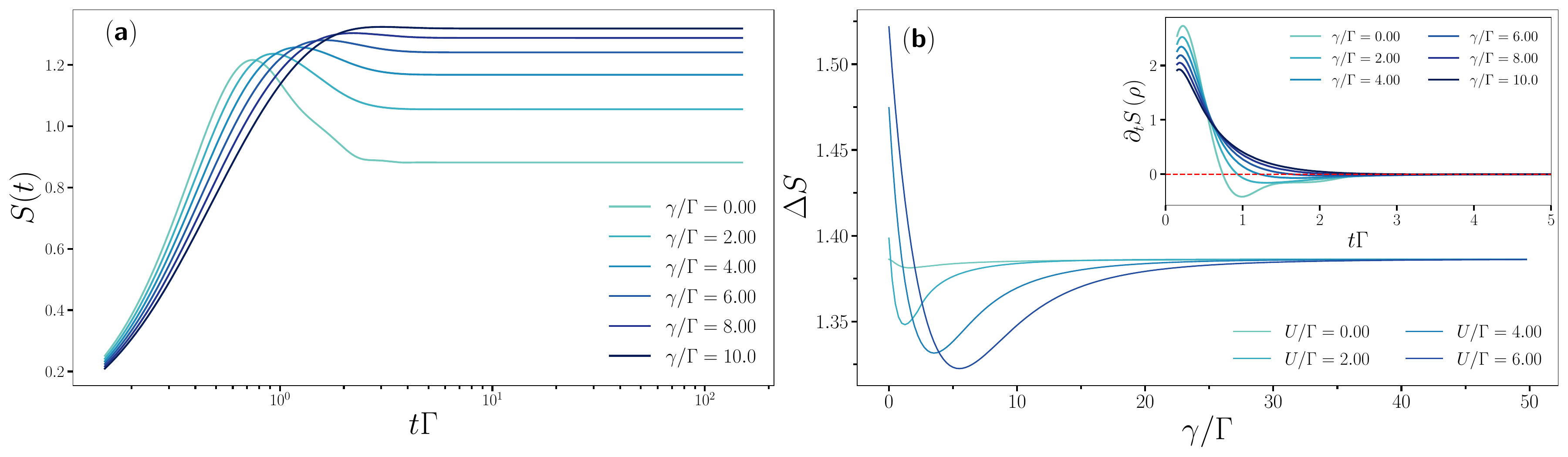}
    \caption{\label{fig:entanglement} Impurity Entanglement Entropy Dynamics - (a) Dynamics of the impurity entropy after tracing out the metallic bath. For small dissipation we see a clear maximum which is then suppressed for increasing dissipation, when the entanglement entropy is monotonously growing towards the steady-state. The crossover between the two regimes is again controlled by a scale of the Zeno type, as we show in the panel (b). We first show the total entanglement rate, which displays a nero of negative values for small $\gamma$. The total change in entropy between the maximum value and the steady-state has a minimum at a value of dissipation $\gamma\sim U  $.    $\Delta S = 2 \max(S) - S_{SS}$
    }
\end{figure*}

\subsection{Impurity Entanglement Entropy} 

We  consider the dynamics of the impurity entanglement entropy, obtained by tracing out the metallic bath and computing the thermal entropy of the reduced density matrix of the system, i.e.
\begin{align}
    S(t)= - \mbox{Tr}\left(\rho_t \mbox{log}\rho_t \right)
\end{align}
where $\rho_t$ is the reduced density matrix of the system after tracing out the metallic bath. This can be directly obtained from the NCA approach which works directly with the evolution operator of the reduced system, i.e. $\rho_t=\mathcal{V}(t,0)\rho_0$.
We emphasize therefore that the state of the system is mixed to begin with, due to the dephasing, therefore the entropy of entanglement also takes contribution from the thermal entropy. In Fig.~\ref{fig:entanglement} we plot the dynamics of the entropy as a function of time for different values of the dissipation rate $\gamma$. In the unitary case, $\gamma=0$, we observe a maximum at short times followed by an approach to a steady-state value. As we include dissipation the maximum remains well visible for small and moderate values of $\gamma$, while for large dissipation the dynamics of the entanglement entropy is monotonous in time towards the steady-state. Interestingly, the crossover between the two regimes of entanglement dynamics is also controlled by the Zeno scale.  To appreciate this point we plot the quantity $\Delta S=2\mbox{max}(S)-S_{SS}$, which for a monotonously growing entropy dynamics reduces to the steady-state value, while in the general case quantify the level of non-monotonicity in $S(t)$, and it is related to the time-integral of the absolute value of the entanglement entropy rate $\partial_t S$. We see that $\Delta S$ has a clear minimum as a function of dissipation, for $\gamma_*\sim U$ as expected from the Zeno effect.

\begin{figure*}[!t] 
 \includegraphics[width=0.95\textwidth]{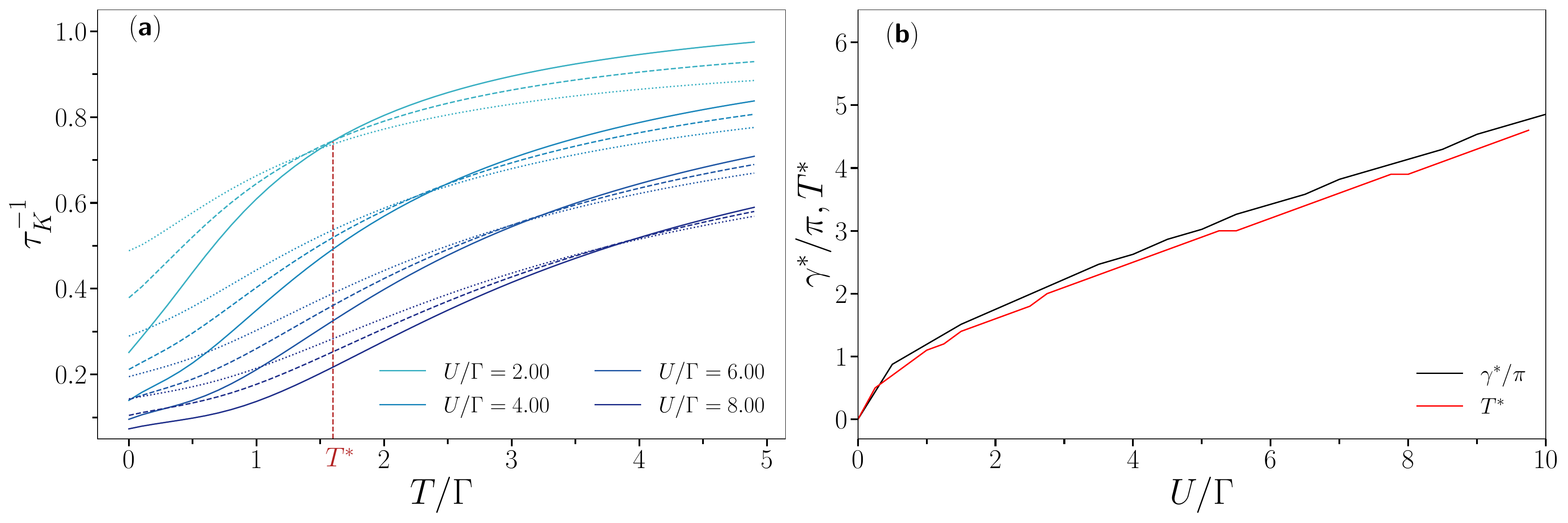}
    \caption{\label{fig:finiteT_vs_U} Interaction dependence of the Spin Relaxation Rate  - (a) Spin relaxation rate $\tau_K^{-1}$ as a function of temperature for different values of $U$ and different values of $\gamma/ \Gamma =0,1,2$ (dashed lines). We see that the data show a crossover around $T^*$, where in the low temperature regime the relaxation rate increases with $\gamma$ while for $T>T^*$ it decreases. As we see in panel (b) the crossover temperature depends on interaction $U$ and its dependence matches closely the dissipation scale $\gamma^*$ controlling the Kondo-Zeno crossover (see Figure 2b of the main text).    }
\end{figure*}

\subsection{Spin Relaxation Rate at Finite Temperature}

Here we provide additional results on the spin relaxation rate $\tau_K^{-1}$ at finite temperature, obtained as discussed in the main text from the real-time dynamics of the magnetization and its decay to zero. In Fig.~\ref{fig:finiteT_vs_U} we plot the spin relaxation rate as a function of temperature (as shown in Fig. 3 of the main text), for respectively different values of $U$ and different values of $\gamma$ (dashed lines). As already mentioned in the main text this quantity displays a crossover in temperature, between a low-$T$ regime where the relaxation rate increases with $\gamma$ and a high-$T$ regime where the rate decreases. As we see in this panel and more clearly in panel (b) of the same figure, the crossover temperature $T^*$ depends strongly on the interaction $U$, increasing linearly with $U$ in the strongly correlated regime. Interestingly this dependence resembles closely the one of the dissipation scale $\gamma^*$ that controls the Kondo-Zeno crossover, that we plot for comparison in panel (b) of the same figure. Overall these results confirm that the temperature dependence of the spin relaxation rate allows to directly access the Kondo-Zeno crossover.

\section{Supplementary Note 4: Schrieffer-Wolff Transformation}

In this section, we provide a detailed derivation of the non-Hermitian Kondo model by using a generalized Schrieffer-Wolff (SW) transformation for open quantum systems~\cite{kessler2012generalized,rosso2020dissipative,nakagawa2020dynamical}, formulated here directly in the vectorized superfermion representation. We start from the Linbdladian in the superfermion representation that we write as
\begin{align}
    \m{L}= \m{L}_0  + \m{L}_{\rm hyb}
\end{align}
where $\m{L}_0$ is the diagonal part (with respect to the hybridization) of the Lindbladian that includes the decoupled impurity and bath terms as well as the local charge dephasing $\m{L}_0=\m{L}_{\rm dot}+\m{L}_{\rm bath}$, while $\m{L}_{\rm hyb}=-i\left(H_{\rm hyb}-\tilde{H}_{\rm hyb}\right)$ is the off-diagonal term.  As in the equilibrium case, the SW transformation allows us to integrate out the bath-impurity coupling terms by introducing a non-unitary (similarity) transformation $S$ such that the new Lindbladian 
\begin{align}
    \mathcal{L}_{\rm eff}=e^S\mathcal{L}e^{-S}=\m{L} + \left[S,\m{L} \right] +\frac{1}{2} \left[S, \left[S,\m{L} \right] \right] + \cdots
\end{align}
is diagonal order by order in a perturbative expansion. In particular this can be achieved by choosing the generator $S$ to be fully off-diagonal with respect to the hybridization and to satisfy the condition
\begin{align}\label{eqn:gen_S}
 \m{L}_{\rm hyb}+[S,\m{L}_0]=0   
\end{align}
which then gives an expression for the effective Lindbladian which reads 
\begin{align}
   \mathcal{L}_{\rm eff}= \m{L}_0+\frac{1}{2}[S,\m{L}_{\rm hyb}]
\end{align}
In the following we first derive the form of the generator $S$ of the Schrieffer-Wolff transformation and then write down the resulting effective Lindbladian operator.

\subsection{The Generator}

The condition for the generator suggests that $S$ should have a structure both diagonal and off-diagonal in the two Hilbert spaces. We therefore parametrize it as following
\begin{align}\label{eqn:S_new}
S=\sum_{\bb{k}\sigma}X_{\bb{k}\sigma}\left(c^{\dagger}_{\bb{k}\sigma}d_{\sigma}-hc\right)+\sum_{\bb{k}\sigma}\tilde{X}_{\bb{k}\sigma}\left(\tilde{c}^{\dagger}_{\bb{k}\sigma}\tilde{d}_{\sigma}-hc\right)
\end{align}
where the operators $X_{\bb{k}\sigma}$ read 
\begin{align}
X_{\bb{k}\sigma}&=s_\bb{k} + t_\bb{k} n_{\bar{\sigma}} + x_{\bb{k}}\tilde{n}+
w_\bb{k}n_{\bar{\sigma}}\tilde{n}+h_{\bb{k}}
\tilde{n}_{\uparrow}\tilde{n}_{\downarrow}+
g_{\bb{k}}n_{\bar{\sigma}} \tilde{n}_{\uparrow}\tilde{n}_{\downarrow}\\
&=\left(s_\bb{k}+t_\bb{k} n_{\bar{\sigma}} \right)+
\left(x_\bb{k}+w_\bb{k} n_{\bar{\sigma}} \right)\tilde{n}+
\left(h_\bb{k}+g_\bb{k} n_{\bar{\sigma}} \right)
\tilde{n}_{\uparrow}\tilde{n}_{\downarrow}
\end{align}
and similarly for $\tilde{X}_{\bb{k}}$ upon switching tilde and not-tilde,i.e.
\begin{align}
\tilde{X}_{\bb{k}\sigma}&=\tilde{s}_\bb{k} + \tilde{t}_\bb{k} \tilde{n}_{\bar{\sigma}} + \tilde{x}_{\bb{k}}n+
\tilde{w}_\bb{k}\tilde{n}_{\bar{\sigma}}n
+\tilde{h}_{\bb{k}}
n_{\uparrow}n_{\downarrow}+
\tilde{g}_{\bb{k}}\tilde{n}_{\bar{\sigma}} n_{\uparrow}n_{\downarrow}\\
&=\left(\tilde{s}_\bb{k}+\tilde{t}_\bb{k}\tilde{n}_{\bar{\sigma}} \right)+
\left(\tilde{x}_\bb{k}+\tilde{w}_\bb{k} \tilde{n}_{\bar{\sigma}} \right)n+
\left(\tilde{h}_\bb{k}+\tilde{g}_\bb{k} \tilde{n}_{\bar{\sigma}} \right)
n_{\uparrow}n_{\downarrow}
\end{align}
While the first two terms in $X_{\bb{k}\sigma},\tilde{X}_{\bb{k}\sigma}$ are diagonal in the Hilbert spaces, and represent straightforward generalisations of the generator for the standard AIM, the last terms represent the new addition due to dissipation. These terms couple the two sectors $\m{H}$ and $\tilde{\m{H}}$ and they are needed to satisfy the condition~(\ref{eqn:gen_S}) due to the structure of the dissipation. To proceed it is convenient to evaluate few commutators between $S$ and different operators, namely
\begin{align}
[S,n_{\sigma}]&=\sum_{\bb{k}}X_{\bb{k}\sigma}\left(c^{\dagger}_{\bb{k}\sigma}d_{\sigma}+hc\right)\\
[S,n_{\uparrow}n_{\downarrow}]&=
\sum_{\bb{k}\sigma}X_{\bb{k}\sigma}\left(c^{\dagger}_{\bb{k}\sigma}d_{\sigma}+hc\right)n_{\bar{\sigma}}\\
[S,n_{\bb{k}\sigma}]&=-X_{\bb{k}\sigma}
\left(c^{\dagger}_{\bb{k}\sigma}d_{\sigma}+hc\right)
\end{align}
Similar relations, upon sending tilde-operators into non-tilde and viceversa, hold for the commutator with $\tilde{n}_{\sigma},\tilde{n}_{\uparrow}\tilde{n}_{\downarrow},\tilde{n}_{\bb{k}\sigma}$. Finally we need to evaluate the commutator between $S$ and the dissipation term coupling the two sectors, namely
\begin{align}
[S,n\tilde{n}]&=
\sum_{\bb{k}\sigma}X_{\bb{k}\sigma}\left(c^{\dagger}_{\bb{k}\sigma}d_{\sigma}+hc\right)\tilde{n}+
\sum_{\bb{k}\sigma}\tilde{X}_{\bb{k}\sigma}\left(\tilde{c}^{\dagger}_{\bb{k}\sigma}\tilde{d}_{\sigma}+hc\right)n
\end{align}
Using these results we can now evaluate the commutator between $S$ and $\mathcal{L}_0$ and obtain the conditions that fix the generator $S$. From this equation, using the commutators above, we get the following conditions 
\begin{align}
\left(\varepsilon_{\bb{k}}-\varepsilon_d+i\gamma/2
\right)X_{\bb{k}\sigma}-(U-i\gamma)X_{\bb{k}\sigma}n_{\bar{\sigma}}-i\gamma X_{\bb{k}\sigma} \tilde{n}&=V_{\bb{k}}\\
\left(\varepsilon_{\bb{k}}-\varepsilon_d-i\gamma/2
\right)\tilde{X}_{\bb{k}\sigma}-(U+i\gamma)\tilde{X}_{\bb{k}\sigma}\tilde{n}_{\bar{\sigma}}+i\gamma \tilde{X}_{\bb{k}\sigma} n&=V_{\bb{k}}
\end{align}
which we can solve to obtain the coefficients
\begin{align}
s_\bb{k}&=\frac{V_\bb{k}}{\left(\varepsilon_{\bb{k}}-\varepsilon_d+i\gamma/2
\right)}\\
t_\bb{k}&=\frac{(U-i\gamma)s_\bb{k}}{\left(\varepsilon_{\bb{k}}-\varepsilon_d-U+3i\gamma/2
\right)}=\frac{(U-i\gamma)V_{\bb{k}}}{\left(\varepsilon_{\bb{k}}-\varepsilon_d-U+3i\gamma/2
\right)\left(\varepsilon_{\bb{k}}-\varepsilon_d+i\gamma/2
\right)}
\\
x_\bb{k}&=\frac{i\gamma s_\bb{k}}{\left(\varepsilon_{\bb{k}}-\varepsilon_d-i\gamma/2
\right)}=\frac{i\gamma V_{\bb{k}} }{\left(\varepsilon_{\bb{k}}-\varepsilon_d\right)^2+\gamma^2/4}
\\
w_\bb{k}&=\frac{(U-i\gamma) x_\bb{k}+i\gamma t_\bb{k}}{\left(\varepsilon_{\bb{k}}-\varepsilon_d-U+i\gamma/2
\right)}
%=\frac{2\gamma^2 }{\left(\varepsilon_{\bb{k}}-\varepsilon_d\right)^2+\gamma^2/4}\frac{(U-i\gamma)V_{\bb{k}}}{\left(\varepsilon_{\bb{k}}-\varepsilon_d+3i\gamma/2\right)\left(\varepsilon_{\bb{k}}-\varepsilon_d+i\gamma/2\right)}=\frac{2\gamma^2 }{\left(\varepsilon_{\bb{k}}-\varepsilon_d\right)^2+\gamma^2/4}t_\bb{k}
\\
h_\bb{k}&=\frac{2i\gamma x_\bb{k}}{\left(\varepsilon_{\bb{k}}-\varepsilon_d-i3\gamma/2
\right)}=-\frac{2\gamma^2 }{\left(\varepsilon_{\bb{k}}-\varepsilon_d\right)^2+\gamma^2/4}\frac{V_\bb{k}}{\left(\varepsilon_{\bb{k}}-\varepsilon_d+i3\gamma/2
\right)}
\\
g_\bb{k}&=\frac{(U-i\gamma) h_\bb{k}+2i\gamma w_\bb{k}}{\left(\varepsilon_{\bb{k}}-\varepsilon_d-U-i\gamma/2
\right)}\\
\end{align}
Similarly, solving for the coefficients entering in the tilde-spce part $\tilde{X}_{\bb{k}\sigma}$ we discover that $\tilde{s}_\bb{k}=s^*_{\bb{k}}$ and similarly for the other coefficients.

\subsection{Derivation of the Effective Lindbladian }

The effective Lindbladian is obtained from the expression
\begin{align}\label{eqn:Leff}
   \mathcal{L}_{\rm eff}= \m{L}_0+\frac{1}{2}[S,\m{L}_{\rm hyb}]=\m{L}_0+\frac{1}{2}[S,\m{L}_{\rm hyb}]=\m{L}_0-\frac{i}{2}[S,H_{\rm hyb}-\tilde{H}_{\rm hyb}]
\end{align}
We now evaluate each of the two commutator separately, involving $H_{\rm hyb},\tilde{H}_{\rm hyb}$ respectively. First we have
%To evaluate the commutator it is convenient to write $\m{L}_{\rm hyb}=-i\left(H_{\rm hyb}-\tilde{H}_{\rm hyb}\right)$ and to consider separately each contribution. We evaluate the first
\begin{align}
[S,H_{\rm hyb}]
&=\sum_{\bb{k}\bb{p}}\sum_{\sigma\sigma'}
X_{\bb{k}\sigma} V_{\bb{p}}[\left(c^{\dagger}_{\textbf{k}\sigma}d_{\sigma}-hc\right),\left(c^{\dagger}_{\textbf{p}\sigma'}d_{\sigma'}+hc\right)]\\
&+\sum_{\bb{k}\bb{p}}\sum_{\sigma\sigma'}
V_{\bb{p}}[X_{\bb{k}\sigma},\left(c^{\dagger}_{\textbf{p}\sigma'}d_{\sigma'}+hc\right)]\left(c^{\dagger}_{\textbf{k}\sigma}d_{\sigma}-hc\right)\\
&+\sum_{\bb{k}\bb{p}}\sum_{\sigma\sigma'}
V_{\bb{p}}[\tilde{X}_{\bb{k}\sigma},\left(c^{\dagger}_{\textbf{p}\sigma'}d_{\sigma'}+hc\right)]\left(\tilde{c}^{\dagger}_{\textbf{k}\sigma}\tilde{d}_{\sigma}-hc\right)
\end{align}
Similarly, for the tilde term we have
\begin{align}
[S,\tilde{H}_{\rm hyb}]
&=\sum_{\bb{k}\bb{p}}\sum_{\sigma\sigma'}
\tilde{X}_{\bb{k}\sigma} V_{\bb{p}}[\left(\tilde{c}^{\dagger}_{\textbf{k}\sigma}\tilde{d}_{\sigma}-hc\right),\left(\tilde{c}^{\dagger}_{\textbf{p}\sigma'}\tilde{d}_{\sigma'}+hc\right)]\\
&+\sum_{\bb{k}\bb{p}}\sum_{\sigma\sigma'}
V_{\bb{p}}[\tilde{X}_{\bb{k}\sigma},\left(\tilde{c}^{\dagger}_{\textbf{p}\sigma'}\tilde{d}_{\sigma'}+hc\right)]\left(\tilde{c}^{\dagger}_{\textbf{k}\sigma}\tilde{d}_{\sigma}-hc\right)\\
&+\sum_{\bb{k}\bb{p}}\sum_{\sigma\sigma'}
V_{\bb{p}}
[X_{\bb{k}\sigma},\left(\tilde{c}^{\dagger}_{\textbf{p}\sigma'}\tilde{d}_{\sigma'}+hc\right)]
\left(c^{\dagger}_{\textbf{k}\sigma}d_{\sigma}-hc\right)
\end{align}
We see that we have therefore to evaluate two types of commutators:
\begin{align}
[\left(c^{\dagger}_{\textbf{k}\sigma}d_{\sigma}-hc\right),\left(c^{\dagger}_{\textbf{p}\sigma'}d_{\sigma'}+hc\right)]&=[c^{\dagger}_{\textbf{k}\sigma}d_{\sigma},
d^{\dagger}_{\sigma'}c_{\textbf{p}\sigma'}]-
[d^{\dagger}_{\sigma}c_{\textbf{k}\sigma},
c^{\dagger}_{\textbf{p}\sigma'}d_{\sigma'}
]\\
&=(c^{\dagger}_{\textbf{k}\sigma}c_{\textbf{p}\sigma'}\delta_{\sigma\sigma'}+hc)-\delta_{\bb{k}\bb{p}}\delta_{\sigma\sigma'}(d^{\dagger}_{\sigma'}d_{\sigma}+hc)
\end{align}
and equivalently for the tilde space
\begin{align}
[\left(\tilde{c}^{\dagger}_{\textbf{k}\sigma}\tilde{d}_{\sigma}-hc\right),\left(\tilde{c}^{\dagger}_{\textbf{p}\sigma'}\tilde{d}_{\sigma'}+hc\right)]=(\tilde{c}^{\dagger}_{\textbf{k}\sigma}\tilde{c}_{\textbf{p}\sigma'}\delta_{\sigma\sigma'}+hc)-\delta_{\bb{k}\bb{p}}\delta_{\sigma\sigma'}(\tilde{d}^{\dagger}_{\sigma'}\tilde{d}_{\sigma}+hc)
\end{align}
as well as 
\begin{align}
[X_{\bb{k}\sigma},\left(c^{\dagger}_{\textbf{p}\sigma'}d_{\sigma'}+hc\right)]=
\left(t_{\bb{k}}+w_{\bb{k}}\tilde{n}+
g_{\bb{k}}\tilde{n}_{\uparrow}\tilde{n}_{\downarrow}\right)\delta_{\bar{\sigma}\sigma^{'}}
\left(c^{\dagger}_{\textbf{p}\sigma'}d_{\sigma'}-hc\right)
\\
[\tilde{X}_{\bb{k}\sigma},\left(\tilde{c}^{\dagger}_{\textbf{p}\sigma'}\tilde{d}_{\sigma'}+hc\right)]=
\left(\tilde{t}_{\bb{k}}+\tilde{w}_{\bb{k}}n+
\tilde{g}_{\bb{k}}n_{\uparrow}n_{\downarrow}\right)\delta_{\bar{\sigma}\sigma^{'}}
\left(c^{\dagger}_{\textbf{p}\sigma'}d_{\sigma'}-hc\right)
\\
[X_{\bb{k}\sigma},\left(\tilde{c}^{\dagger}_{\textbf{p}\sigma'}\tilde{d}_{\sigma'}+hc\right)]=
-\left(x_{\bb{k}}+w_{\bb{k}}n_{\bar{\sigma}}\right)\left(\tilde{c}^{\dagger}_{\textbf{p}\sigma'}\tilde{d}_{\sigma'}-hc\right)\\
+\left(h_{\bb{k}}+g_{\bb{k}}n_{\bar{\sigma}}\right)
\left(\tilde{c}^{\dagger}_{\textbf{p}\sigma'}\tilde{d}_{\sigma'}-hc\right)
\left(\delta_{\sigma^{'}\downarrow}\tilde{n}_{\uparrow}+\delta_{\sigma^{'}\uparrow}\tilde{n}_{\downarrow}\right)\\
[\tilde{X}_{\bb{k}\sigma},\left(c^{\dagger}_{\textbf{p}\sigma'}d_{\sigma'}+hc\right)]=
-\left(\tilde{x}_{\bb{k}}+\tilde{w}_{\bb{k}}\tilde{n}_{\bar{\sigma}}\right)\left(c^{\dagger}_{\textbf{p}\sigma'}d_{\sigma'}-hc\right)\\
+\left(\tilde{h}_{\bb{k}}+\tilde{g}_{\bb{k}}\tilde{n}_{\bar{\sigma}}\right)
\left(c^{\dagger}_{\textbf{p}\sigma'}d_{\sigma'}-hc\right)
\left(\delta_{\sigma^{'}\downarrow}\tilde{n}_{\uparrow}+\delta_{\sigma^{'}\uparrow}\tilde{n}_{\downarrow}\right)
\end{align}
We can then evaluate the commutators between $S$ and $H_{\rm hyb},\tilde{H}_{\rm hyb}$ entering Eq.~(\ref{eqn:Leff}). We obtain
\begin{align}
[S,H_{\rm hyb}]
&=\sum_{\bb{k}\bb{p}\sigma}X_{\bb{k}\sigma} V_{\bb{p}}\left(
c^{\dagger}_{\textbf{k}\sigma}c_{\textbf{p}\sigma}+hc\right)-2\sum_{\bb{k}\sigma}
X_{\bb{k}\sigma} V_{\bb{k}}d^{\dagger}_{\sigma}d_{\sigma}
\\
&-\sum_{\bb{k}\bb{p}\sigma}
V_{\bb{p}}\left(t_{\bb{k}}+w_{\bb{k}}\tilde{n}
+g_{\bb{k}}\tilde{n}_{\uparrow}\tilde{n}_{\downarrow}\right)\left(c^{\dagger}_{\textbf{p}\bar{\sigma}}d_{\bar{\sigma}}-hc\right)\left(c^{\dagger}_{\textbf{k}\sigma}d_{\sigma}-hc\right)\\
&-\sum_{\bb{k}\bb{p}}\sum_{\sigma\sigma'}
V_{\bb{p}}
\left(\tilde{x}_{\bb{k}}+\tilde{w}_{\bb{k}}\tilde{n}_{\bar{\sigma}}\right)
\left(c^{\dagger}_{\textbf{p}\sigma'}d_{\sigma'}-hc\right)\left(\tilde{c}^{\dagger}_{\textbf{k}\sigma}\tilde{d}_{\sigma}-hc\right)\\
&+\sum_{\bb{k}\bb{p}}\sum_{\sigma\sigma'}
V_{\bb{p}}
\left(\tilde{h}_{\bb{k}}+\tilde{g}_{\bb{k}}\tilde{n}_{\bar{\sigma}}\right)
\left(c^{\dagger}_{\textbf{p}\sigma'}d_{\sigma'}-hc\right)n_{\bar{\sigma'}}\left(\tilde{c}^{\dagger}_{\textbf{k}\sigma}\tilde{d}_{\sigma}-hc\right)
\end{align}
while for the one involving $\tilde{H}_{\rm hyb}$ is is sufficient to swap all tilde operators into non-tilde ones in the equation above and take complex conjugation of the coefficients. Using these results and Eq.~(\ref{eqn:Leff}) we can rewrite the effective Lindbladian in a more compact form. To this extent it is useful to introduce the spinors $\Phi^{\dagger}_\bb{k}=\left( c^{\dagger}_{\bb{k}\uparrow}c^{\dagger}_{\bb{k}\downarrow}\right)$ and 
$\Psi^{\dagger}_d=\left( d^{\dagger}_{\uparrow} d^{\dagger}_{\downarrow}\right)$. The effective Lindbladian is composed of different contributions, 
\begin{align}\label{eqn:Leff_final}
       \m{L}_{\rm eff} &= \m{L}_0 + \m{L}_{\rm bath}+\m{L}_{\rm Kondo}+
       \m{L}_{\rm scatt}+\m{L}_{\rm pair}+\m{L}_{\rm diss}
\end{align}
which we now discuss in detail. $\m{L}_{\rm Kondo}$ describes a Kondo coupling between the dot spin and the spin of the bath, 
\begin{align}
\m{L}_{\rm Kondo}=  i\sum_{\bb{k}\bb{q}} J_{\bb{q}\bb{k}} \left( \Phi_\bb{q}^\dagger \frac{\Vec{\sigma}}{2} \Phi_\bb{k}\right) \cdot \Vec{S}_d   + \text{TildeVersion}
\end{align}
with a coupling 
$$J_{\bb{q}\bb{k}}  = V_\bb{q} \left( t_{\bb{k}} +w_{\bb{k}} \tilde{n}
     \right) + V_\bb{k} \left( t_{\bb{q}} +w_{\bb{q}}
    \tilde{n} \right)$$
which is in general complex. $\m{L}_{\rm scatt}$ describes a scattering potential for the conduction electrons, 
\begin{align}
\m{L}_{\rm scatt}= -i \sum_{\bb{k}\bb{q}} \left( \frac{1}{2}W_{\bb{q}\bb{k}} + \frac{1}{4} J_{\bb{q}\bb{k}} \left( \Psi_d^\dagger \Psi_d \right) \right) \Phi_\bb{q}^\dagger \Phi_\bb{k}   + \text{TildeVersion}
\end{align}
with coupling constant
$$
W_{\bb{q}\bb{k}} = 
    V_\bb{q} \left( s_{\bb{k}} +x_{\bb{k}} \tilde{n}
     %\left[ \Tilde{n}_\uparrow + \Tilde{n}_\downarrow \right]
     \right) + V_\bb{k} \left( s_{\bb{q}} +x_{\bb{q}}%\left[ \Tilde{n}_\uparrow + \Tilde{n}_\downarrow \right]
    \tilde{n} \right)
$$
$\m{L}_{\rm pair}$ describes a pair tunneling term 
\begin{align}
\m{L}_{\rm pair}=  \frac{i}{2} \sum_{\bb{q}\bb{k}\sigma} T_{\bb{q}\bb{k}}\left( c_{\bb{q}\sigma}^\dagger c_{\bb{k}\bar{\sigma}}^\dagger d_{\bar{\sigma}} d_\sigma + h.c\right) + \text{TildeVersion}
\end{align}
with amplitude 
$$
T_{\bb{q}\bb{k}} =V_\bb{k}\left[ t_\bb{k} + g_\bb{k} \Tilde{n}_\uparrow \Tilde{n}_\downarrow + w_\bb{k} \tilde{n}\right]
$$
$\m{L}_{\rm dot}$ describes a local Lindblad contribution
%\begin{align}
%\m{L}_{\rm dot}= 
%-  \sum_\bb{k} V_\bb{k}\left( 2 t_\bb{k} + 2 w_\bb{k} - \Tilde{h}_\bb{k}\right) n_\uparrow n_\downarrow -  \sum_{\bb{k}} V_\bb{k} \left( h_\bb{k} - 2\Tilde{t}_\bb{k} - 2 \Tilde{w}_\bb{k}\right) \Tilde{n}_\uparrow \Tilde{n}_\downarrow - 2 \sum_{\bb{k}} V_\bb{k} \left( g_\bb{k} - \Tilde{g}_\bb{k} \right) n_\uparrow n_\downarrow \Tilde{n}_\uparrow \Tilde{n}_\downarrow 
%\end{align}
\begin{align}\label{eqn:Ldot}
\m{L}_{\rm dot}=&  
 i \sum_\bb{k} V_\bb{k}\left( 2 t_\bb{k} + 2 w_\bb{k}\Tilde{n} + 2 g_\bb{k} \Tilde{n}_\uparrow \Tilde{n}_\downarrow \right) n_\uparrow n_\downarrow - i  \sum_{\bb{k}} V_\bb{k} \left( 2\Tilde{t}_\bb{k} + 2 \Tilde{w}_\bb{k} n + 2 \Tilde{g}_\bb{k} n_\uparrow n_\downarrow \right) \Tilde{n}_\uparrow \Tilde{n}_\downarrow \notag \\ & + i \left( \sum_{\bb{k}}V_\bb{k} \left[ s_\bb{k} + x_\bb{k} \Tilde{n} + h_\bb{k} \Tilde{n}_\uparrow \Tilde{n}_\downarrow \right] \right) \left( \sum_\sigma  n_\sigma \right) -i \left( \sum_\bb{k} V_\bb{k} \left[ \Tilde{s}_\bb{k} + \Tilde{x}_\bb{k} n + \Tilde{h}_\bb{k} n_\uparrow n_\downarrow \right] \right) \left( \sum_\sigma \Tilde{n}_\sigma \right)
\end{align}
and finally $\m{L}_{\rm diss}$ describes a dissipative term 
\begin{align}
\m{L}_{\rm diss}&= 
 i\left[\sum_{\bb{q}\bb{k}} H_{\bb{q}\bb{k}} \left( \Phi_\bb{q}^\dagger \frac{\Vec{\sigma}}{2} \Phi_\bb{k}\right) \cdot \Vec{S}_d  + \sum_{\bb{q}\bb{k}} \left( \frac{1}{2}I_{\bb{q}\bb{k}} + \frac{1}{4} H_{\bb{q}\bb{k}} \left( \Psi_d^\dagger \Psi_d \right) \right) \Phi_\bb{q}^\dagger \Phi_\bb{k} \right] \Tilde{n}_\uparrow 
 \Tilde{n}_\downarrow + \text{TildeVersion}\\&
  +\frac{i}{2} \sum_{\bb{q}\bb{k}\sigma \mu}\Gamma_{\bb{q}\bb{k}} \left( c_{\bb{k}\sigma}^\dagger d_\sigma + c_{\bb{k}\sigma} d_\sigma^\dagger \right)\left( \Tilde{c}_{\bb{q}\mu}^\dagger \Tilde{d}_\mu + \Tilde{c}_{\bb{q}\mu} \Tilde{d}_\mu^\dagger \right) 
  %\left[ x_\bb{k}V_\bb{q} - \Tilde{x}_\bb{q} V_\bb{k}\right] \notag \\&
  %+ \frac{1}{2} \sum_{\bb{q}\bb{k}\sigma \mu} \left( c_{\bb{k}\sigma}^\dagger d_\sigma + c_{\bb{k}\sigma} d_\sigma^\dagger \right)\left( \Tilde{c}_{\bb{q}\mu}^\dagger \Tilde{d}_\mu + \Tilde{c}_{\bb{q}\mu} \Tilde{d}_\mu^\dagger \right) \Tilde{n}_{\bar{\mu}} \left[h_\bb{k} V_\bb{q} - \Tilde{w}_\bb{q} V_\bb{k} \right] \notag \\ &+
   %\frac{1}{2} \sum_{\bb{q}\bb{k}\sigma \mu} \left( c_{\bb{k}\sigma}^\dagger d_\sigma + c_{\bb{k}\sigma} d_\sigma^\dagger \right)\left( \Tilde{c}_{\bb{q}\mu}^\dagger \Tilde{d}_\mu + \Tilde{c}_{\bb{q}\mu} \Tilde{d}_\mu^\dagger \right) \Tilde{n}_{\bar{\mu}} n_{\bar{\sigma}} \left[ g_\bb{k} V_\bb{q} - \Tilde{g}_\bb{q} V_\bb{k} 
  %\right]
\end{align}
with coupling constants 
\begin{align}
    & I_{\bb{q}\bb{k}} = V_\bb{q} h_\bb{k} +  V_\bb{k} h_\bb{q} \notag \\ & 
    H_{\bb{q}\bb{k}} = V_\bb{q} g_\bb{k} +  V_\bb{k} g_\bb{q} \notag \\ &
    \Gamma_{\bb{q}\bb{k}}=\left[ x_\bb{k}V_\bb{q} - \Tilde{x}_\bb{q} V_\bb{k}\right]+\Tilde{n}_{\bar{\mu}}\left[h_\bb{k} V_\bb{q} - \Tilde{w}_\bb{q} V_\bb{k} \right] +\Tilde{n}_{\bar{\mu}} n_{\bar{\sigma}} 
    \left[ g_\bb{k} V_\bb{q} - \Tilde{g}_\bb{q} V_\bb{k} 
  \right]
\end{align}

It is first of all useful to comment on the structure of the effective Lindbladian as compared to the equilibrium case, when the Schrieffer-Wolff on the Anderson Impurity model leads to a Kondo model. Indeed we notice that some of the effective terms which are generated in Eq.~(\ref{eqn:Leff_final}) would be also present  in equilibrium. This is the case for example of the Kondo coupling $\m{L}_{\rm Kondo}$, the scattering potential $\m{L}_{\rm scatt}$, the pair-hopping term $\m{L}_{\rm pair}$ or the local term $\m{L}_{\rm dot}$. The effect of dissipation here is to renormalize the coupling constants $J_{\bb{q}\bb{k}},W_{\bb{q}\bb{k}},T_{\bb{q}\bb{k}}$, leading to either complex (dissipative) interactions or to introduce coupling between Hilbert space sectors (tilde/non-tilde operators), which as we know describe dissipative processes. For example, we can see in Eq.~(\ref{eqn:Ldot}) that dissipation induces a renormalization to the bare Hubbard repulsion $U$, compatible with the spectral function shown in the main text, which shows a drift of the Hubbard bands with $\gamma$, as well as an effective doublon-doublon dissipation.

Dephasing is however also responsible for generating intrinsically new terms such as $\m{L}_{\rm diss}$, which couples operators in the two Hilbert spaces and introduce new dissipative processes which for example create or destroy a doublon/holon with a rate $\Gamma_{\bb{q}\bb{k}}$. These dissipative terms couple the singly occupied spin sector to the charge fluctuations and therefore allow the system to go out from the singly occupied manifold. In the equilibrium low-energy limit of the AIM, one then proceeds by projecting the effective theory in the singly-occupied manifold, which is the low-energy one, to obtain a Kondo model. In the next section we show that the dynamics in this subspace is described by a non-Hermitian Kondo model. Our analysis on the effective Lindbladian, as well as our NCA results, show however that this projection can only be valid in certain regimes of dephasing, roughly for $\gamma<U$, before heating due to doublon production take over.

%We see several different terms entering and it is useful to comment on their importance and significance. The first line describe the effective spin-spin (Kondo) interaction and the scattering potential. These terms control the dynamics in the singly occupied manifold and we note that they are diagonal in the index controlling the Hylbert space. In the next section we will interpret those terms as an effective non-Hermitian Kondo model. We then recognize the pair-hopping term and the renormalization to the local dot Hamiltonian. Finally the last four lines describe terms which allows to go out of the singly occupied manifold by creating doublons-holons pair. Importantly these terms are all off-diagonal in the Hilbert space index, i.e. they describe genuine dissipation and they are ultimately responsible for the doublon production at long times. However for small values of dephasing these term do not play a role since the system remains close to the singly occupied subspace. In the next section we show that the dynamics in this subspace is described by a non-Hermitian Kondo model.

\begin{figure*}[!ht] 
 \includegraphics[width=0.95\textwidth]{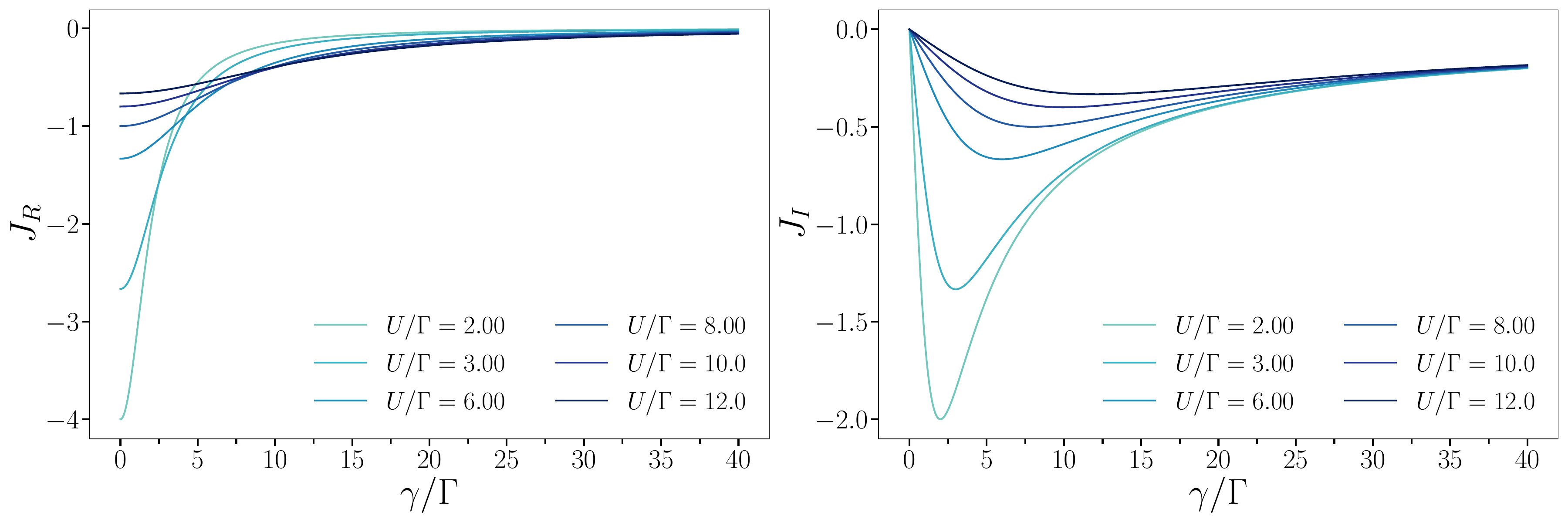}
    \caption{\label{fig:SpinExchangeCoupling} 
    Spin-exchange complex coupling in the Half filling case $\epsilon_d = -U/2$ and at the Fermi level $\epsilon_\bb{k}=0$ - (Left Panel) Real part of Kondo coupling  (Right Panel) imaginary part of the Kondo coupling  }
\end{figure*}

\subsection{Effective Linbdladian and non-Hermitian Kondo model}

We now discuss the structure of the effective Lindbladian $\m{L}_{\rm eff}$ in the singly occupied half-filled sector. To this extent we introduce projectors for the degrees of freedom in each Hilbert space $\m{H}/ \Tilde{\m{H}}$ 
\begin{align}
    P &= \sum_{\sigma} n_\sigma (1- n_{\bar{\sigma}})  \quad \text{and} \quad Q=1-P\\
    \tilde{P} &= \sum_{\sigma} \tilde{n}_\sigma (1- \tilde{n}_{\bar{\sigma}})  \quad \text{and} \quad \tilde{Q}=1-\tilde{P}   
\end{align}
with $P^2 =P$ and $\left[P , \Tilde{P} \right] = \left[ P,\Tilde{Q}\right] =0$. We note that $P$ projects on the dot single occupied states (either spin up or down) while $Q$ on the empty/double occupied states, and similarly for $\tilde{P},\tilde{Q}$ in $\tilde{\m{H}}$. The effective Lindbladian reads
\begin{align}
    \m{L}_{\rm eff}&=  \left(\Tilde{P}+ \Tilde{Q}\right) \left(P+Q\right) \m{L}_{eff}\left(P+Q\right)\left(\Tilde{P}+ \Tilde{Q}\right) 
\end{align}
Since the total Lindbladian of the system is particle hole symmetric, the low energy sector in the symmetric case is characterized by $\Psi_d^\dagger \Psi_d=1 $ and $\epsilon_d = -U /2$. In this limit many of the terms entering $\m{L}_{eff}$ cancel out and we obtain
\begin{align}
    \m{L}_{\rm eff} = \m{L}_{\rm bath} -i \left( H_{\rm eff} - \Tilde{H}_{\rm eff} \right) 
\end{align}
where the $H_{\rm eff}$ describes a non-Hermitian Kondo model with a potential scattering term
\begin{align}
    H_{\rm eff} = -\sum_{\bb{k}\bb{q}} J_{\bb{q}\bb{k}} \left( \Phi_\bb{q}^\dagger \frac{\Vec{\sigma}}{2} \Phi_\bb{k}\right) \cdot \Vec{S}_d + \sum_{\bb{k}\bb{q}} \left( \frac{1}{2}W_{\bb{q}\bb{k}} + \frac{1}{4} J_{\bb{q}\bb{k}} \left( \Psi_d^\dagger \Psi_d \right) \right) \Phi_\bb{q}^\dagger \Phi_\bb{k}
\end{align}
where the complex spin-exchange coupling and the scattering potential are given by,
\begin{align}
    &J_{\bb{q}\bb{k}}  = V_\bb{q} \left( t_{\bb{k}} +w_{\bb{k}} \right) + V_\bb{k} \left( t_{\bb{q}} +w_{\bb{q}} \right) =
    \left(\frac{V_\bb{q}V_\bb{k}(U-i\gamma)}{\left(\varepsilon_\bb{k}-\varepsilon_d-U+i\gamma/2\right)\left(\varepsilon_\bb{k}-\varepsilon_d-i\gamma/2\right)}+
    \frac{V_\bb{q}V_\bb{k}(U-i\gamma)}{\left(\varepsilon_\bb{q}-\varepsilon_d-U+i\gamma/2\right)\left(\varepsilon_\bb{q}-\varepsilon_d-i\gamma/2\right)}
    \right)
    \notag \\ & W_{\bb{q}\bb{k}} = V_\bb{q} \left( s_{\bb{k}} +x_{\bb{k}} \right) + V_\bb{k} \left( s_{\bb{q}} +x_{\bb{q}} \right)=
    \left(\frac{V_\bb{q}V_\bb{k}}{\varepsilon_\bb{k}-\varepsilon_d-i\gamma/2}+
    \frac{V_\bb{q}V_\bb{k}}{\varepsilon_\bb{q}-\varepsilon_d-i\gamma/2}
    \right)
\end{align}
If one evaluate the Kondo coupling $J_{\bb{q}\bb{k}}$ for bath momenta at the Fermi energy, i.e. $\epsilon_\bb{k} = \epsilon_\bb{q} = 0$, which is the relevant limit at long-times and for $\varepsilon_d=-U/2$ as required by particle-hole symmetry  the expressions of the couplings become

Moreover the complex-valued spin-exchange coupling constant $J$ at the Fermi level is obtained by setting $\epsilon_\bb{k} = \epsilon_\bb{q} = 0$  as well as $\varepsilon_d=-U/2$ in the equation above , this approximation is valid because only the physics close to the Fermi energy is studied
\begin{align}
    J_{\bb{q}\bb{k}} &= J_R+iJ_I =
    -\frac{8V^2}{(U-i\gamma)}\\
     W_{\bb{q}\bb{k}} &= W =\frac{4V^2}{(U-i\gamma)}
\end{align}
Moreover, in this limit, the potential scaterring terms cancel out and the dissipative part of $\m{L}_0$ does not contribute to the low energy physics. The behavior of the real and imaginary part of the Kondo coupling is plotted in Fig.~\ref{fig:SpinExchangeCoupling} showing that the real-part is quickly suppressed with $\gamma$ while the imaginary part $J_I$ displays the characteristic Zeno crossover around $\gamma\sim U$.

\section{Supplementary Note 5: Spin versus Charge Monitoring}

\begin{figure*}[t!]
 	\centering
    \includegraphics[width=\textwidth]{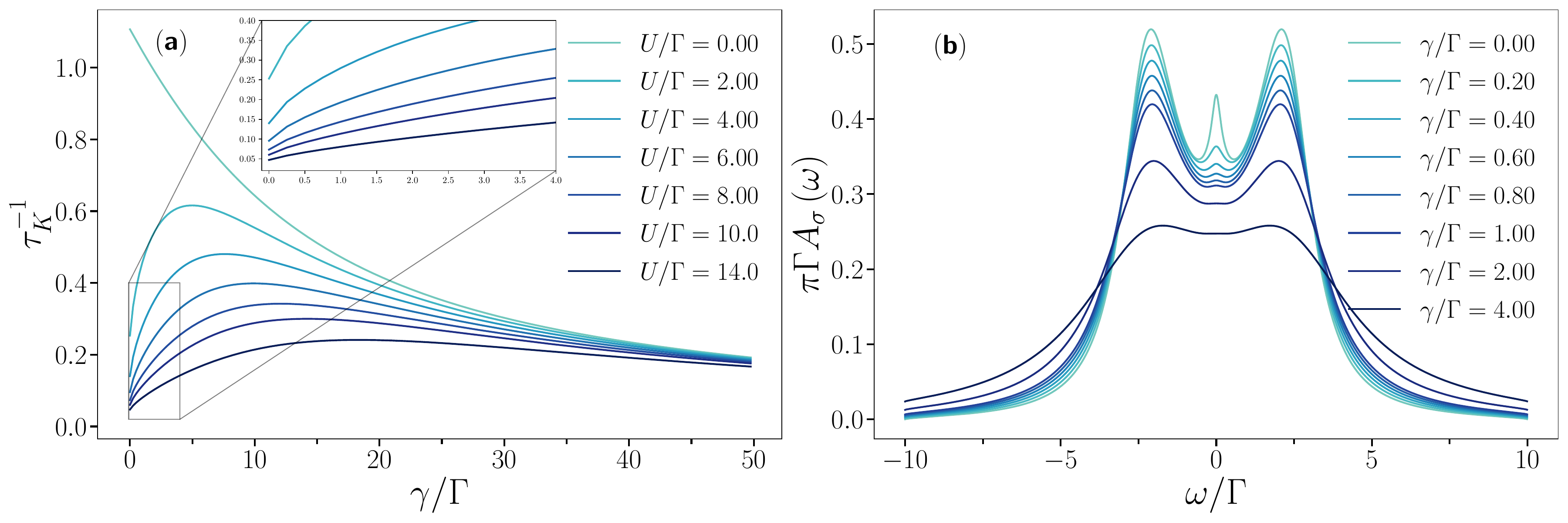}
    \caption{\label{fig:SpinDephasing}  Dissipative Anderson Impurity Model - (a) Decay rate $\tau_K^{-1}$ in function of the dissipation $\gamma$. (b) Dot spectral function for the half-filled dissipative AIM and for $U = -2 \epsilon_d = 4 \Gamma $ and increasing values of spin dephasing $\gamma$. Here the dissipation considered is the charge dephasing $L=  S_z$. }
\end{figure*}

In the previous Sections we have argued that the problem of an Anderson dot under continuous monitoring of its total charge displays an interesting non-Hermitian Kondo physics, due to the fact that ultimately the monitoring of the total charge tries to quench charge fluctuations but leaves unaffected the spin sector. To provide further evidence in support of this picture in this Section we focus on a dissipative Anderson model with continuous monitoring of the dot spin. This problem is described again by a Linbdlad master equation of the form
\begin{align}\label{eqn:lindblad}
    \partial_t \rho_t = -i \left[ H, \rho_t \right] + \gamma \left( L \rho_t L^\dagger -\frac{1}{2} \{ L^\dagger L , \rho_t \}\right)
\end{align}
where $H$ is the Anderson impurity Hamiltonian given in the main text
\begin{align}\label{eq:H_aim}
H=\sum_{\bb{k},\sigma}\varepsilon_{\bb{k}}c^\dagger_{\bb{k},\sigma}c_{\bb{k},\sigma}+ \sum_{\bb{k},\sigma} \left(V_{\bb{k}} d_{\sigma}^\dagger c_{\bb{k},\sigma} + h.c \right)+H_{\rm dot}\,.
\end{align}
while the jump operator is now proportional to the dot's spin $z$ component, i.e. it reads now
\begin{align}
L =L^\dagger =  \sum_{\sigma}\sigma n_{\sigma}\equiv S_z\,,
\end{align}
where $\gamma$ is now the spin monitoring (or spin dephasing) rate. For this model we study again the spin relaxation time and the steady-state dot spectral function, that we plot in Fig.~\ref{fig:SpinDephasing}. The spin dynamics starting from a factorised initial state displays an exponential relaxation towards zero, from which we can extract the lifetime that we plot in Fig.~\ref{fig:SpinDephasing} (panel a). We clearly see that also for the spin dephasing case the Kondo-Zeno crossover is well visible. To understand the effect of spin-dephasing on Kondo physics we have to focus on the regime of weak dissipation. Here we see that the spin relaxation time appears to grow linearly with $\gamma$, even in the strongly correlated regime $U\gg\Gamma$, differently from the case of charge monitoring. This suggests that spin dephasing has a stronger effect on Kondo physics. This result is further confirmed by looking at the dot spectral function that we plot in Fig.~\ref{fig:SpinDephasing} (panel b). Here we can see that quite strikingly the coherent Kondo peak is suppressed very effectively by spin depahsing, which on the other hand leave almost untouched the Hubbard bands.
This result confirms the picture that monitoring the charge of the dot provides a more gentle way to control and modify the Kondo physics rather than acting directly on the spin degrees of freedom. 

We can obtain further insights onto this problem by turning again to the Schrieffer-Wolff transformation discussed in the Supplementary Note 4. In particular we can now repeat the previous analysis and derive an effective model for the spin-dephasing problem. This reads
\begin{align}
   \mathcal{L}_{\rm eff}=e^S\mathcal{L}e^{-S}\simeq  \m{L}_0+\frac{1}{2}[S,\m{L}_{\rm hyb}]
\end{align}
where the genetator $S$ takes now the form
\begin{align}\label{eqn:S_new}
S=\sum_{\bb{k}\sigma}X_{\bb{k}\sigma}\left(c^{\dagger}_{\bb{k}\sigma}d_{\sigma}-hc\right)+\sum_{\bb{k}\sigma}\tilde{X}_{\bb{k}\sigma}\left(\tilde{c}^{\dagger}_{\bb{k}\sigma}\tilde{d}_{\sigma}-hc\right)
\end{align}
where the operators $X_{\bb{k}\sigma}$ read 
\begin{align}
    X_{\bb{k}\sigma}&=s_{\bb{k}\sigma} + t_{\bb{k}\sigma} n_{\bar{\sigma}} + x_{\bb{k}\sigma} \tilde{S}_z + w_{\bb{k}\sigma} n_{\bar{\sigma}} \tilde{S}_z
    + h_{\bb{k}\sigma} \tilde{S}_z^2 + g_{\bb{k}\sigma} n_{\bar{\sigma}}\tilde{S}_z^2\,.
\end{align}
We note that the $z-$component of the dot spin appears now explicitly in the generator (together with its equivalent in the tilde space). This is crucial in order to be able to fix the coefficients of the generator $S$ via the condition $\m{L}_{\rm hyb}+[S,\m{L}_0]=0$. After some lengthy algebra one finally for the effective Lindbladian the form
\begin{align}
    \m{L}_{eff} = \m{L}_0 + \m{L}_{\rm Kondo} + \m{L}_{\rm scatt} + \m{L}_{\rm pair}+\m{L}_{\rm Kondo-diss}
\end{align}
where $\m{L}_{0}$ describes the diagonal part of the Lindbladian, $\m{L}_{\rm Kondo},\m{L}_{\rm scatt}, \m{L}_{\rm pair}$ represent respectively the Kondo coupling, the potential scattering term and the dissipative pair-hoppings discussed already in the charge monitoring case while $\m{L}_{Kondo-diss}$ is a genuinely new term that arises in the spin-dephasing case. It takes the form
\begin{align}
\m{L}_{\rm Kondo-diss}=
i\sum_{\bb{q}\bb{k}}J^{zz}_{\bb{k}\bb{q}}\tilde{S}_z (\Phi_\bb{q}^\dagger \sigma^z  \Phi_\bb{k}) +\mbox{Tilde Version} 
\end{align}
namely a dissipative Kondo coupling between the $z-$ component of the dot and bath magnetization. Crucially, this term survives even in the single occupied sector and describe a genuine dissipative term, coupling the two Hilbert spaces (Keldysh contour), thus driving the system away from the physics of the non-Hermitian Kondo model present in the charge monitoring problem. We can speculate that it is precisely this term the one responsible for the fragility of the Kondo peak in presence of spin dephasing as shown in the NCA results.  A full theoretical study of the Anderson Impurity Model with spin dephasing is beyond the scope of this work and left for future study.

\end{document}